\documentclass[12pt]{article}
\usepackage{amssymb}
\usepackage{amsmath}

\newcommand{\CC}{{\mathbb C}}
\newcommand{\RR}{{\mathbb R}}
\newcommand{\ZZ}{{\mathbb Z}}
\newcommand{\PP}{{\mathbb P}}

\renewcommand{\SS}{{\mathbb S}}
\newcommand{\End}{{\rm End}}
\newcommand{\Ker}{{\rm Ker}}
\newcommand{\Tr}{{\rm Tr}}
\newcommand{\diag}{{\rm diag}}
\newcommand{\ra}{\rightarrow}
\newcommand{\hE}{{\hat E}}
\newcommand{\hA}{{\hat A}}
\newcommand{\hphi}{{\hat \phi}}
\newcommand{\hX}{{\hat X}}

\newcommand{\bz}{\bar{z}}
\newcommand{\bs}{\bar{s}}

\newcommand{\hD}{\hat{D}}

\newcommand{\ha}{\hat{a}}
\newcommand{\hP}{\hat{P}}

\newcommand{\bpartial}{\bar{\partial}}

\newcommand{\ind}{{\rm Ind}}

\newcommand{\hDd}{\hat{D}^{\dagger}}
\newcommand{\ot}{\otimes}
\newcommand{\bD}{{\bar{\cal D}}}

\newcommand{\cK}{{\mathcal K}}
\newcommand{\cM}{{\mathcal M}}

\newcommand{\hSS}{{\hat \SS}}

\newcommand{\ups}{{\mathfrak v}}
\newcommand{\bX}{{\bar X}}

\newcommand{\Dd}{D^{\dagger}}

\newcommand{\C}{{\bf C}}

\newcommand{\tA}{{\tilde A}}
\newcommand{\Sc}{{\bf S}}
\renewcommand{\S}{S}
\newcommand{\bsl}{{\backslash}}
\newcommand{\op}{{\oplus}}

\newcommand{\nn}{\nonumber}

\newcommand{\phih}{\hat{\phi}}

\newcommand{\N}{{\cal N}}
\newcommand{\rank}{{\rm rank}}

\newcommand{\nb}{{\bf NB}}
\newcommand{\bk}{{\bar k}}
\newcommand{\uth}{\breve{\theta}}

\newcommand{\tr}{{\rm tr}}

\newcommand{\talpha}{{\tilde\alpha}}
\newcommand{\rre}{{\rm Re}}
\newcommand{\iim}{{\rm Im}}

\title{\bf  Periodic Monopoles With Singularities\\ And\\ {\cal N}=2 Super-QCD}
\author{
Sergey A. Cherkis\thanks{email: cherkis@physics.ucla.edu}\\
\it TEP, UCLA Physics Department, Los Angeles, CA 90095-1547
 \rm
\and 
Anton Kapustin\thanks{email: kapustin@ias.edu}\\
\it Institute for Advanced Study, Olden Lane, Princeton, NJ 08540
}

\begin{document}

\begin{titlepage}

\renewcommand{\thepage}{ }

\maketitle

\begin{sloppypar} 
\begin{abstract} 
We study solutions of the Bogomolny equation on $\RR^2\times\SS^1$
with prescribed singularities. We show that Nahm transform establishes
a one-to-one correspondence between such solutions and solutions of the 
Hitchin equations on a punctured cylinder with the eigenvalues of the Higgs field 
growing at infinity in a particular manner. The moduli spaces
of solutions have natural hyperk\"ahler metrics of a novel kind. We show that
these metrics describe the quantum Coulomb branch of certain ${\cal N}=2$ $d=4$ 
supersymmetric gauge theories on $\RR^3\times\SS^1$.
The Coulomb branches of the corresponding uncompactified theories have been 
previously determined by E.~Witten using the M-theory fivebrane. We show that
the Seiberg-Witten curves of these theories are identical to the spectral
curves associated to solutions of the Bogomolny equation on $\RR^2\times\SS^1$.
In particular, this allows us to rederive Witten's results without recourse 
to the M-theory fivebrane.
 
\end{abstract} 

\vspace{-8.5in}

\parbox{\linewidth}
{\small\hfill \shortstack{IASSNS-HEP-00/82\\ UCLA-00TEP-25 \\ CITUSC/00-58}}

\end{sloppypar}
\end{titlepage}

\pagestyle{headings}

\tableofcontents

\section{Introduction}

Let $X$ be an oriented
Riemannian 3-manifold, $E$ a unitary vector bundle on $X$, $A$ a connection on $E$,
and $\phi$ a Hermitian section of $\End(E)$. The Bogomolny equation is a nonlinear
differential equation
\begin{equation}
F_A=*d\phi.
\end{equation}
In the case $X\cong\RR^3$ with the standard metric, solutions of this equation
have been extensively studied from many different viewpoints~(see~\cite{AH}
and references therein).

In~\cite{usone} we studied solutions of the Bogomolny equation for $\rank(E)=2$ 
and $X=\RR^2\times\SS^1$ with a standard metric. The norm of the Higgs field 
was assumed to grow logarithmically at infinity. Such solutions were called
periodic monopoles.
They are topologically classified by a positive integer, the monopole charge.
Using Nahm transform, we showed that there is a one-to-one correspondence between
periodic monopoles with charge $k$ and solutions of rank $k$ Hitchin equations
on a cylinder with a particular asymptotic behavior. Hitchin equations are the
dimensional reduction of the self-duality equation to two dimensions. 

In this paper we study solutions of the Bogomolny equation on 
$\RR^2\times\SS^1$ with $n$ points deleted. 
The behavior of $A,\phi$ in the neighborhood of a deleted point is that of 
a Dirac monopole minimally embedded in the nonabelian gauge group. 
The eigenvalues of the
Higgs field $\phi$ are allowed to grow logarithmically at infinity. 
Solutions of this kind
will be called periodic monopoles with $n$ singularities.
We mostly deal with 
the case $\rank(E)=2$,
but we also sketch how our results can be generalized to higher rank. 
For $\rank(E)=2$ periodic monopoles with $n$ singularities are topologically
classified by a single integer $k$ which satisfies $2k\geq n$. We call
this integer the monopole charge.

There are several reasons to study periodic monopoles with singularities. First
of all, their moduli spaces carry natural hyperk\"ahler metrics of a novel kind.
For example, for $k=2$ and $n=4$ the centered moduli space is
a smooth four-dimensional hyperk\"ahler manifold with a distinguished complex
structure. As a complex manifold it is isomorphic to a blow-up of $(\CC\times S)/\ZZ_2$ at four
points, where $S$ is an elliptic curve, and $\ZZ_2$ acts by reflection on $\CC$ and by
a nontrivial element of $Aut(S)$ on $S$. The four blown-up points are the fixed
points of the $\ZZ_2$ action. This complex
manifold is a noncompact analogue of the Kummer surface.
It can be argued that the natural hyperk\"ahler metric on it
is complete, nondegenerate, and asymptotically locally flat. For a fascinating introduction
to noncompact approximations to K3 metrics see~\cite{HitchinK3}.

{}From the physical point of view, these moduli spaces are interesting because they
provide exact low-energy effective actions for $N=2$ $d=4$ gauge theories
compactified on a circle. ``Exact'' here  means that both perturbative and 
non-perturbative quantum corrections are included. While the effective
action of $N=2$ $d=4$ gauge theories on $\RR^4$ can be computed by a variety of 
methods~\cite{SW,Witten,geomeng,probe}, the analogous problem on $\RR^3\times\SS^1$ remained 
intractable so far. The main reason is the necessity to sum over an infinite number of 
instanton contributions,
including virtual BPS monopoles wrapping $\SS^1$. In our previous paper~\cite{usone}
we explained how the moduli space of periodic monopoles can be used to solve this problem
in the case of $N=2$ Yang-Mills theory without matter. Periodic monopoles with
singularities allow one to solve $N=2$ gauge theories with gauge group $SU(k)$
and matter in the fundamental representation. 

Finally, studying periodic monopoles with singularities provides a new example of 
Nahm transform, which is a differential-geometric analogue of the Fourier-Mukai
transform. The Bogomolny equation is a reduction of the self-duality equation
to three dimensions. In general, Nahm transform maps a solution of the former equation
into a solution of some different system of equations, which is also a reduction of the
self-duality equation. The precise form of this new system of equations depends on 
the boundary conditions imposed on the Bogomolny equation. 
For example, Nahm transform takes monopoles on $\RR^3$ with finite energy 
to solutions of the so-called Nahm equations, which are the reduction of the 
self-duality equation to one dimension. Periodic monopoles without singularities
are mapped to solutions of the Hitchin equations on a cylinder~\cite{usone}.
We will see that Nahm transform establishes a one-to-one correspondence between
periodic monopoles with singularities and solutions of Hitchin equations on a cylinder
with singularities. 

The singularities of the Hitchin data on the cylinder are so-called tame singularities.
Solutions of Hitchin equations on compact curves with such singularities were
previously studied by C.~Simpson~\cite{Simp} and others.
Since Hitchin equations are conformally-invariant, one may also wish to compactify
the cylinder to a $\PP^1$ by adding two points at infinity. For general $k$ and $n$
the singularities of the Hitchin data at the two added points are not tame.
Nevertheless, it appears that the Hitchin-Kobayashi correspondence, if properly 
understood, continues to hold in this situation. It would be very interesting to 
understand this issue in detail. 

There is one special situation $(2k=n)$ where all four singularities of the Hitchin data on 
$\PP^1$ are tame. In this case the rank of the Hitchin data is $k.$ In particular, if
we want the moduli space of the Hitchin data to be the noncompact Kummer surface mentioned
above, we have to set $k=2$ and consider rank-two Hitchin equations on $\PP^1$ with four 
tame singularities. This moduli space can be reinterpreted as the centered moduli space of
two periodic monopoles with four Dirac-type singularities.

Monopoles on $\RR^3$ 
with Dirac-type singularities have been studied in~\cite{usrthree3}, 
\cite{usrthree2}, and \cite{usrthree1}.
Their moduli spaces are asymptotically locally flat
hyperk\"ahler manifolds which can be used to solve $N=4$ $d=3$ gauge theories with 
matter. The present work can be viewed as an extension
of both~\cite{usrthree3} and~\cite{usone}.

In this paper we explain the relation between periodic monopoles with singularities
and $N=2$ gauge theories, and study their Nahm transform. The properties of the
moduli space will be explored in a forthcoming publication~\cite{ALG}.

\section{Periodic Monopoles With Singularities}\label{sec:U2mon}

\subsection{Periodic $U(2)$ Monopoles With Singularities}\label{U2}
In this section we give the precise definition of a periodic monopole with 
singularities.
Let $X$ be $(\RR^2\times \SS^1)\backslash\{p_1,\ldots,p_n\}$, where
$p_i,i=1,\ldots,n,$ are distinct points. We will parametrize $\SS^1$ by
$\chi\in \RR/(2\pi\ZZ)$ and $z\in\CC\cong\RR^2$. Consider a $U(2)$ bundle $E$ on $X$.
Its topological type is completely determined by $n$ integers $e_1,\ldots,e_n$,
the values of the first Chern class of $E$ on small 2-spheres surrounding the
points $p_1,\ldots,p_n$. We will assume that $e_i=\pm 1$ for all $i$.

Let us set $\phi_0(r)=-1/(2r)$.
Let us define a $U(1)$ connection $A_0(x)$ on a line bundle on $\RR^3\backslash\{0\}$ 
by $dA_0=*d\phi_0$. The first Chern class of this line bundle evaluated on a 2-sphere
enclosing the origin is one. 

Let $\phi_\infty(z,\chi)=\frac{\log|z|}{2\pi}$ be a function on 
$\cM=(\RR^2\times\SS^1)\backslash\{z=0\}$. We cover $\cM$ with two coordinate patches,
$U_0=\{\arg z\neq \pi\}$, and $U_1=\{\arg z\neq 0\}$, where $\arg z$ is assumed to
take values in the interval $(-\pi,\pi]$. Let $L$ be a unitary line bundle on $\cM$ 
with the following transition function between $U_0$ and $U_1$:
$$
g(z,\chi)=\begin{cases} 
1, & \iim z<0 \\
e^{-i\chi}, & \iim z>0.
\end{cases}
$$
The first Chern class of this line bundle evaluated on any 2-torus of the form $|z|=const$
is one.
We define a unitary connection on $L$ by
\begin{equation}\label{helm}
A_\infty=\begin{cases}
\frac{\arg z}{2\pi}d\chi, & \arg z\neq \pi,\\
\frac{\arg (-z)-\pi}{2\pi}d\chi, & \arg z\neq 0.
\end{cases}
\end{equation}
The connection $A_\infty$ satisfies $dA_\infty=*d\phi_\infty$.

A periodic monopole on $E$ is a solution of the Bogomolny equation such that the
connection and the Higgs field behave as
\begin{align}\label{singHiggs}
\phi(x)&\sim g_i(x)\begin{pmatrix} e_i\phi_0(r_i) & 0 \\ 0 & 0\end{pmatrix}
g_i(x)^{-1}+O(1),\\ \label{dersingHiggs}
d_A\phi(x)&\sim g_i(x)\begin{pmatrix} e_i d\phi_0(r_i) & 0 \\ 0 & 0 \end{pmatrix}
g_i(x)^{-1}+O(1),\\ \label{singA}
A(x)&\sim g_i(x) \begin{pmatrix} e_iA_0(x-x_i) & 0 \\ 0 & 0\end{pmatrix}
g_i(x)^{-1}+ ig_i(x) dg_i(x)^{-1}+O(1),
\end{align}
near the $i^{\rm th}$ singularity (here $r_i$ is the distance to the
$i^{\rm th}$ singularity and $g_i(x)$ is a $U(2)$-valued function), while at 
infinity their behavior is given by
\begin{align}\label{infHiggs}
2\pi\phi(x)\sim & g(x)\diag\left(2\pi\ell_1\phi_\infty+v_1+
\rre\frac{\mu_1}{z}, 2\pi\ell_2\phi_\infty+v_2+\rre\frac{\mu_2}{z}\right)g(x)^{-1}\nn\\
& +O\left(\frac{1}{|z|^2}\right),\\ 
2\pi d_A \phi(x)\sim & g(x)\diag\left(2\pi\ell_1 d\phi_\infty - \rre\frac{\mu_1 dz}{z^2},
2\pi\ell_2 d\phi_\infty - \rre\frac{\mu_2 dz}{z^2}\right)g(x)^{-1}\nn\\ \label{derinfHiggs}
& + O\left(\frac{1}{|z|^3}\right),\\ 
2\pi A(x)\sim & g(x)\diag\left(2\pi\ell_1 A_\infty(x)+\left[ b_1+
\iim \frac{\mu_1}{z}\right]d\chi+\alpha_1 d\arg z,\right. \\
& \left. 2\pi\ell_2 A_\infty(x)+\left[ b_2 + 
\iim \frac{\mu_2}{z}\right]d\chi+\alpha_2 d\arg z\right) g(x)^{-1}
\nn\\
& +2\pi ig(x) dg(x)^{-1}+O\left(\frac{1}{|z|^2}\right). \label{infA}
\end{align}
Here $g(x)$ is a $U(2)$-valued function, $\ell_i,v_i,b_i,\alpha_i\in\RR, i=1,2$,
$\mu_i\in\CC, i=1,2$.
The integers $e_i$ will be referred to as the abelian charges of the periodic 
monopole.
We can assume that $\ell_1\geq \ell_2$ without loss of generality. If $\ell_1=\ell_2$, 
we will require in addition that $v_1>v_2$. Physically, this means that for 
large $|z|$ the $U(2)$ gauge symmetry is broken to 
$U(1)\times U(1)$ by the Higgs field. 

The numbers $\alpha_1$ and $\alpha_2$ are not gauge-invariant: a gauge transformation
may shift them by $2\pi m, m\in\ZZ$. Therefore we prefer to regard $\alpha_{1,2}$
as taking values in $\RR/(2\pi\ZZ).$

{}From these formulas it is easy to see that $\ell_1+\ell_2$ is the value of the 
first Chern class of $E$ on any sufficiently large 2-torus enclosing all the singularities.
Since this 2-torus is homologous to the union of $n$ small 2-spheres surrounding the
singularities, it follows that 
$$
\ell_1+\ell_2=\sum e_i.
$$
Both $\ell_1$ and $\ell_2$ are integers.
Indeed, the eigenvalues of the Higgs field outside a sufficiently
large compact region are distinct, and therefore we can define a line subbundle
of $E$ associated with the largest eigenvalue of the Higgs field. The value
of its Chern class on a large 2-torus is $\ell_1$, therefore $\ell_1$ must be an integer.
Hence $\ell_2$ is also an integer.

There are also relations between the continuous parameters appearing 
in~(\ref{singHiggs}) and (\ref{infHiggs}). If we denote by $(z_i,\chi_i)$
the coordinates of the point $p_i$, $i=1,\ldots,n$, then these relations read:
\begin{align}
\mu_1+\mu_2&=-\sum_{i=1}^n e_i z_i, \\
\alpha_1+\alpha_2&=\sum_{i=1}^n e_i \chi_i.
\end{align}
The derivation of these relations is presented in the next subsection.

We define the nonabelian charge of a monopole to be 
\begin{equation}\nn
k=\frac{1}{2}(\ell_1-\ell_2+n).
\end{equation}
It is easy to see that $k$ is a positive integer; in fact, since 
$\ell_1\geq \ell_2$, it satisfies 
$$
2k\geq n.
$$
The asymptotic behavior of the Higgs field is completely fixed once we specify
$e_i, i=1,\ldots,n,$ and $k$. 

Let $n_\pm$ be the total number of singularities with $e_i=\pm 1$. By definition,
$n_++n_-=n$. The integers
$\ell_1,\ell_2$ which determine the behavior of the Higgs field at infinity can be
expressed in terms of $k,n_+,$ and $n_-$:
$$
\ell_1=k-n_-,\qquad \ell_2=n_+ -k.
$$ 

\subsection{Periodic $SO(3)$ Monopoles With Singularities}

A closely related problem is that of $SO(3)$ monopoles with singularities on 
$\RR^2\times\SS^1$. These are solutions of the Bogomolny equation with traceless
$A$ and $\phi$. The behavior of $A,\phi$ near the singularities is given
by
\begin{align}\label{singSO}
\phi(x)&\sim g_i(x)\begin{pmatrix} \frac{1}{2}\phi_0(r_i)& 0 \\ 0 & 
-\frac{1}{2}\phi_0(r_i)\end{pmatrix}
g_i(x)^{-1}+O(1),\\
d_A\phi(x)&\sim g_i(x)\begin{pmatrix} \frac{1}{2}d\phi_0(r_i)& 0 \\ 0 & 
-\frac{1}{2}d\phi_0(r_i)\end{pmatrix}
g_i(x)^{-1}+O(1),\\
A(x)&\sim g_i(x) \begin{pmatrix} \frac{1}{2}A_0(x-x_i) & 0 \\ 0 & 
-\frac{1}{2}A_0(x-x_i)\end{pmatrix}
g_i(x)^{-1}+ ig_i(x) dg_i(x)^{-1}\nn\\
& +O(1),
\end{align}
where $g_i(x)$ are again $U(2)$-valued functions. The behavior at infinity
is given by
\begin{align}\label{infSO}
4\pi\phi(x)\sim & g(x)\diag\left(2\pi k_\infty\phi_\infty+v+
\rre\frac{\mu}{z}, -2\pi k_\infty\phi_\infty-v-\rre\frac{\mu}{z}\right) g(x)^{-1}\nn\\
& +O\left(\frac{1}{|z|^2}\right),\\
4\pi d_A\phi(x)\sim & g(x)\diag\left(2\pi k_\infty d\phi_\infty - \rre\frac{\mu dz}{z^2},
-2\pi k_\infty\phi_\infty +\rre\frac{\mu dz}{z^2}\right) g(x)^{-1}\nn\\
& +O\left(\frac{1}{|z|^3}\right),\\
4\pi A(x) \sim & g(x)\diag\left(2\pi k_\infty A_\infty(x)+\left[b+\iim\frac{\mu}{z}
\right]d\chi+\alpha d\arg z,\right.\nn\\
& \left. - 2\pi k_\infty A_\infty(x)-\left[b+\iim\frac{\mu}{z}\right]d\chi
-\alpha d\arg z\right)  g(x)^{-1}\nn\\
& + 4\pi ig(x) dg(x)^{-1}+O\left(\frac{1}{|z|^2}\right),
\end{align} 
where $g(x)$ is a $U(2)$-valued function, $k_\infty,v,b,\alpha\in\RR$, $\mu\in\CC$. 
We may assume that $k_\infty\geq 0$ 
without loss of generality. The number $k_\infty$ is, in fact, an integer, as
it measures the value of first Chern class of the eigenbundle 
corresponding to the positive eigenvalue of $\phi$ on a large 2-torus.  

The second Stiefel-Whitney class of this $SO(3)$
bundle evaluated on a small 2-sphere surrounding  the $i^{\rm th}$ singularity
is $1$, therefore it cannot be lifted to an $SU(2)$ bundle. The Stiefel-Whitney
class of the bundle evaluated on a large 2-torus is $k_\infty \mod 2$. Since the
large 2-torus is homologous to the sum of the small 2-spheres surrounding the
singularities, it follows that $k_\infty=n \mod 2$, where $n$ is the total
number of singularities. We will define the
nonabelian charge of an $SO(3)$ monopole to be $(k_\infty+n)/2$. In view of the
above, the nonabelian charge is greater or equal than $n/2.$

The relation between the $U(2)$ and $SO(3)$ periodic monopoles is the following.
If we decompose $U(2)$ monopole fields $A$ and $\phi$ into a trace part and a 
trace-free part,
$$
A=A_{tr}+A_{tf},\qquad \phi=\phi_{tr}+\phi_{tf},
$$
then $(A_{tr},\phi_{tr})$ and $(A_{tf},\phi_{tf})$ separately satisfy the Bogomolny
equation. It is easy to see that the behavior of $(A_{tf},\phi_{tf})$
near the singularities is described by~(\ref{singSO}), while their behavior
at infinity is described by~(\ref{infSO}) with 
$$
k_\infty=\ell_1-\ell_2,\quad
v=v_1-v_2,\quad
b=b_1-b_2,\quad
\mu=\mu_1-\mu_2,\quad
\alpha=\alpha_1-\alpha_2.
$$
Furthermore, $(A_{tr},\phi_{tr})$ represents $n$ periodic Dirac monopoles
and therefore obeys~\cite{usone}
\begin{align}\label{abelHiggs}
\phi_{tr}=&\frac{1}{2\pi} (v_1+v_2)+\frac12 \sum_i e_i V(x-x_i)\sim 
\frac{1}{2\pi}(v_1+v_2)+
\frac12(\ell_1+\ell_2)\frac{\log |z|}{2\pi}\nn\\
 & -\frac{1}{4\pi} \sum_i e_i \rre\frac{z_i}{z}+O\left(\frac{1}{|z|^2}\right), \\ \nn
A_{tr}\sim &\frac12(\ell_1+\ell_2) A_\infty(x)+\frac{1}{4\pi}\left[b_1+b_2-
\sum_i e_i \iim\frac{z_i}{z}\right] d\chi+
\frac12 (\alpha_1+\alpha_2)\frac{ d\arg z}{2\pi}\nn\\
&+O\left(\frac{1}{|z|^2}\right).
\end{align}
Here the function $V(x)$ is given by
$$
\frac{\log(4\pi)-\gamma}{2\pi}-\frac{1}{2}{\sideset{}{'}
\sum_{p=-\infty}^{\infty}} \left[ 
\frac{1}{\sqrt {|z|^2+(\chi-2\pi p)^2}}-\frac{1}{2\pi |p|}\right],
$$
where the prime means that for $p=0$ the second term in the square brackets 
must be omitted, and $\gamma$ is the Euler's constant.
The equation~(\ref{abelHiggs}) implies that $\mu_1$ and $\mu_2$ cannot be 
chosen arbitrarily, but must satisfy
$$
\mu_1+\mu_2=-\sum_{i=1}^n e_i z_i.
$$

There is another important relation constraining the parameters of the periodic 
monopole. It relates the asymptotic parameters $\alpha_i, i=1,2,$ in 
Eq.~(\ref{infHiggs}) and the positions $\chi_1, \chi_2,\ldots,\chi_n$ of the
singularities along the $\SS^1$.
Consider the holonomy of $A_{tr}$ along a circle $|z|=R,\chi=\chi_0$. For
$R\ra\infty$ the holonomy tends to
$$
\frac{1}{2}(\alpha_1+\alpha_2-(l_1+l_2)\chi_0).
$$ 
This follows from Eq.~(\ref{infHiggs}) and the choice of trivialisation specified
by Eq.~(\ref{helm}). On the other hand, it is clear 
from symmetry considerations that the curvature $F_{z\bz}$ constructed
from $A_{tr}$ vanishes identically on the plane $\chi=\chi_0$ if $\chi_0$
is the $\chi$-coordinate of the center-of-mass of the singularities, i.e. if
\begin{equation}
\chi_0=\frac{1}{n}\sum e_i \chi_i.
\end{equation}
By Stockes' theorem, the limiting holonomy of $A_{tr}$ must
vanish for this value of $\chi_0$. This implies that
\begin{equation}\label{shroud}
\alpha_1+\alpha_2=\sum e_i \chi_i.
\end{equation}

To summarize, to any $U(2)$ 
periodic monopole with $n$ singularities one can associate
a $U(1)$ periodic monopole with $n$ singularities and an $SU(2)/\ZZ_2=SO(3)$
periodic monopole with $n$ singularities. The nonabelian charge of the $SO(3)$
monopole is equal to the nonabelian charge of the $U(2)$ monopole. 
Moreover, it is easy to see that the induced map on the moduli spaces of 
$U(2)$ and $SO(3)$
monopoles is an isometry.
This basically follows from the fact that a periodic $U(1)$
monopole is completely determined by $e_i$, the location of the singularities,
and the asymptotics of the Higgs field at infinity, and therefore it has 
no moduli.

We just learned that the moduli
space of periodic $U(2)$ monopoles depends only on the total number
of singularities $n$ and the nonabelian charge $k$, but not on the individual
values of $e_i$. If we were only interested in the moduli space, we could
have set all $e_i$ to be one, for example. We prefer to keep $e_i$ arbitrary,
since the Nahm transform depends on the abelian charges in a nontrivial way.

\subsection{Periodic $U(m)$ Monopoles With Singularities}\label{PUM}

One can similarly define periodic monopoles with singularities for other classical groups.
As an example, we consider $U(m)$ periodic monopoles. The asymptotic behavior
of the Higgs field and the connection at infinity is given by
\begin{align}\label{infHiggsM}
2\pi\phi(x) &\sim  g(x)\diag\left( 2\pi\ell_1 \phi_\infty(x)+v_1+
\rre\frac{\mu_1}{z}, 2\pi\ell_2\phi_\infty(x)+v_2+\rre\frac{\mu_2}{z},\right.\nn\\ 
&\ldots,\left. 2\pi\ell_m\phi_\infty(x)+v_m+\rre\frac{\mu_m}{z}\right) g(x)^{-1}+
O\left(\frac{1}{|z|^2}\right),\\ 
2\pi d_A\phi(x)&\sim g(x)\diag\left( 2\pi\ell_1 d\phi_\infty(x) -
\rre\frac{\mu_1 dz}{z^2}, 2\pi\ell_2d\phi_\infty(x) - \rre\frac{\mu_2 dz}{z^2},\right.\nn\\ 
&\ldots,\left. 2\pi\ell_m d\phi_\infty(x) - \rre\frac{\mu_m dz}{z^2}\right) g(x)^{-1}+
O\left(\frac{1}{|z|^3}\right),\nn\\ \label{infAM}
2\pi A(x)&\sim g(x)\diag\left(2\pi\ell_1 A_\infty(x)+
\left[b_1 +\iim\frac{\mu_1}{z}\right]d\chi+\alpha_1 
d\arg z,\right.\\
& \left. 2\pi\ell_2 A_\infty(x)+\left[b_2+\iim\frac{\mu_2}{z}\right]
d\chi+\alpha_2 d\arg z,\ldots,\right.\nn\\
&\left. 2\pi \ell_m A_\infty(x)+
\left[b_m+\iim\frac{\mu_m}{z}\right]d\chi+
\alpha_m d\arg z\right) g(x)^{-1}\nn\\
&+ 2\pi ig(x) dg(x)^{-1}+O\left(\frac{1}{|z|^2}\right).\nn
\end{align}
We may assume that $\ell_1\geq \ell_2 \geq \ldots \geq \ell_m$. In addition we assume that if $\ell_i=\ell_{i+1}$, then $v_i>v_{i+1}$. 
The numbers $\ell_1,\ldots,\ell_m$ must be integers for $A(x)$ to be a well-defined connection.
One way to see this is to note that since all of the eigenvalues of $\phi$ are distinct for large enough $|z|$, there
is a well-defined splitting of $E$ into the eigenbundles of $\phi$, and $\ell_1,\ldots, \ell_m$ are equal to
the values of the first Chern class of these line bundles on a large 2-torus 
$|z|=const.$ 

The singularities at the points $p_1,\ldots,p_n$ are given by the Dirac monopole minimally embedded into
the $U(m)$ gauge group:
\begin{align}\label{singHiggsM}
\phi(x)&\sim g_i(x)\diag(e_i\phi_0(r_i),0,\ldots,0)g_i(x)^{-1}+O(1),\nn\\ \nn
d_A\phi(x)&\sim g_i(x)\diag(e_i d\phi_0(r_i),0,\ldots,0)g_i(x)^{-1}+O(1),\\ \nn
A(x)&\sim g_i(x)\diag(e_i A_0(x-x_i),0,\ldots,0)g_i(x)^{-1}+ig_i(x)dg_i(x)^{-1}+O(1),
\end{align}
where $e_i=\pm 1$. The first Chern class of $E$ evaluated on a small 2-sphere surrounding the $i^{\rm th}$
singularity is equal to $e_i$. Since the first Chern class of $E$ evaluated on a large 2-torus $|z|=const$
is equal to $\ell_1+\ldots+\ell_m$, we have the relation
$$
\sum_{j=1}^m \ell_j=\sum_{i=1}^n e_i.
$$

We define the nonabelian
charge of a monopole to be a vector $(k_1,k_2,\ldots,k_{m-1})$ with components
$$
k_p=n_- +\sum_{i=1}^p \ell_i.
$$ 
Clearly, the integers $\ell_j$ are completely determined by the abelian charges 
$e_i$ and the nonabelian charge $(k_1,\ldots,k_{m-1})$.

Given a periodic $U(m)$ monopole with singularities, one can decompose its fields into a trace-free and
a trace part which separately satisfy the Bogomolny equation. The trace part is completely determined by
the abelian charges $e_i$. The traceless part defines a $U(m)/U(1)$ periodic monopole with
singularities. To understand the nature of the singularities, recall that a vector bundle with the structure group
$U(m)/U(1)=SU(m)/\ZZ_m$ has a characteristic class with values in $H^2(X,\ZZ_m)$ which generalizes
the second Stiefel-Whitney class. This class measures the obstruction for lifting the $U(m)/U(1)$ bundle to an $SU(m)$
bundle. In the physics literature it is known as the t'Hooft magnetic flux. 
One can show that the
value of the t'Hooft magnetic flux on a small 2-sphere surrounding the $i^{\rm th}$ singularity is equal to $e_i \mod m$.
Thus the $U(m)/U(1)$ bundle corresponding to a $U(m)$ monopole with singularities cannot be lifted to 
an $SU(m)$ bundle. Note that for $m>2$ the singularity with $e_i=+1$ is distinct from $e_i=-1$ even after passing
to traceless fields. 

Conversely, given a periodic $U(m)/U(1)$ monopole with singularities, one can unambiguously reconstruct a
periodic $U(m)$ monopole with singularities. This happens because the trace part, being a periodic $U(1)$ monopole,
with singularities, is completely determined by $e_i$ and the asymptotics at infinity.

As in the case of $U(2)$ monopoles, there are constraints between various continuous
parameters appearing in~(\ref{infHiggsM}) and (\ref{singHiggsM}), namely
\begin{align}
\sum_{j=1}^m \mu_j&=-\sum_{i=1}^n e_i z_i,\\
\sum_{j=1}^m \alpha_j&=\sum_{i=1}^n e_i\chi_i.
\end{align}
These constraints can be derived by computing $A_{tr}$ and $\phi_{tr}$ and
comparing with the known expressions for a $U(1)$ periodic monopole with 
singularities.

%%%%%%%%%%%%%%%%%%%%%%%%%%%%%%%%%%%%%%%%%%%%%%%%%%%%%%%%%%%%%%%%%%%%

\section{$N=2$ Gauge Theories Compactified On A Circle}
\label{gauge} 
In this section we explain the relevance of periodic monopoles with singularities for
understanding quantum properties of supersymmetric gauge theories. 
Chalmers and Hanany~\cite{ChH} were the first to realize that the metric on the moduli
space of certain supersymmetric gauge theories in three dimensions is identical
to the metric on the moduli space of BPS monopoles. This relation was used to great
effect by  many authors, notably by Hanany and Witten~\cite{HW}. Later on, this relation
was extended to four-dimensional $N=2$ gauge theories compactified on a circle of arbitrary
radius $R$~\cite{GS,SavAnton,Anton}, and it was shown that the quantum moduli space of 
many interesting theories
of this kind coincides with the moduli space of self-duality equations or their reductions. Thus a
difficult quantum-mechanical problem can often be converted to a much simpler problem of
studying the moduli space of certain partial differential equations. In particular, in the decompactification
limit $R\ra\infty$ one can recover all the results of Seiberg, Witten, and others on the moduli space
of four-dimensional $N=2$ gauge theories. 

The precise form of the PDE one has to study depends on the gauge theory in question. For example,
certain finite $N=2$ gauge theories (the so-called quiver theories) are solved in terms of
instantons on $\RR^2\times T^2$~\cite{SavAnton,Anton}. $N=2$ super-Yang-Mills with gauge group $SU(k)$ and no
hypermultiplets are solved in terms of monopoles on $\RR^2\times\SS^1$~\cite{usone}. We will see below that
periodic monopoles with $n$ singularities and nonabelian charge $k$ are relevant for $N=2$
$SU(k)$ gauge theory with $n$ massive hypermultiplets.

\subsection{The Geometry Of The Coulomb Branch}

Consider an $N=2$ $SU(k)$ gauge theory in a generic vacuum on the Coulomb branch, where
the expectation value of the Higgs field in the vector multiplet breaks the gauge
group down to its maximal torus. The low-energy effective theory is described by
$k-1$ abelian vector multiplets which contain $k-1$ complex scalars $\phi$, $k-1$
photons $A$, and $2(k-1)$ Mayorana fermions. Thus the moduli space of the theory
is a $k-1$-dimensional complex manifold. $N=2$ supersymmetry requires the
metric on the moduli space to be a special K\"ahler metric.

Now consider compactifying the theory on a circle of radius $R$. At length scales larger
than $R$ the theory is effectively three-dimensional. Its bosonic fields include
$k-1$ complex scalars, $k-1$ periodic real scalars originating from Wilson lines of
the four-dimensional photons along the compactified direction, and $k-1$ periodic
real scalars obtained by dualizing $k-1$ three-dimensional photons. All in all,
the moduli space of the effective three-dimensional theory is $4(k-1)$-dimensional.
Its metric is required to be hyperk\"ahler by supersymmetry. 

Far from the origin of the Coulomb branch, the metric can be found by first flowing
to the infrared in the four-dimensional theory, and then dimensionally reducing on
a circle. This is possible because the low-energy effective theory in four-dimensions
is free, and thus no renormalization group flow occurs upon compactification.
The resulting moduli space is fibered over the moduli space of the four-dimensional
theory by $2(k-1)$-dimensional tori. The metric on the fibers is flat, and thus the
metric on the total space far along the Coulomb brancg has a $U(1)^{2k-2}$ 
isometry~\cite{SW3,Anton}.

As one moves towards the origin of the Coulomb branch, the form of the metric
starts to deviate from this simple form. In particular, while the four-dimensional instantons
respect the the $U(1)^{2k-2}$ isometry, the Euclidean BPS monopoles wrapping
the compactified direction do not. These effects are exponentially small far along
the Coulomb branch, but are very important near the origin. They tend to smooth out the
singularities of the naive metric obtain by dimensional reduction.

{}From the above discussion it is clear that the asymptotic behavior of the metric
on the moduli space of the compactified theory is determined by the four-dimensional
physics alone. If the four-dimensional theory is asymptotically free, or finite, the
metric on the moduli space is locally flat at infinity. For the $SU(k)$ gauge
theory with $n$ hypermultiplets this happens if $2k\geq n$. The interpretation
of this restriction in terms of periodic monopoles will be explained below.

\subsection{String Theory Picture}

The relation between periodic monopoles and $N=2$ gauge theories emerges if one embeds 
these gauge theories into string theory in a particular way, which we now explain.

$N=2$ $SU(k)$ gauge theories can be realized in IIA string theory by suspending
$k$ D4-branes between two parallel NS5-branes. We shall
assume that the NS5-branes' world-volume is along the $0,1,2,3,4,5$ directions,
and their positions in the $7,8,9$ directions coincide. The NS5-brane with
smaller (resp. larger) $x^6$ coordinate will be called the left (resp. right) NS5-brane.
The $k$ D4-branes
are infinite in the $0,1,2,3$ directions and span a finite interval in the $6$
direction. The two boundaries of the D4-brane worldvolume lie on the NS5-branes.
The direction $3$ will be assumed to be periodic with period $2\pi R$. 

\begin{center}
\begin{tabular}{c|cccccccccc}
 &0&1&2&\raisebox{1pt}{$\bigcirc$}\hspace{-8.6pt}3&4&5&6&7&8&9\\
\hline
NS5&x&x&x&x&x&x& & & & \\
D4&x&x&x&x& & &x& & &
\end{tabular}
\end{center}

The world-volume theory on the D4-branes reduces in the infrared limit to the
$\N=2$ $SU(k)$ Yang-Mills theory on $\RR^3\times \SS^1$, where $x^0,x^1,x^2$ are affine
coordinates on $\RR^3$ and $x^3$ parametrizes $\SS^1$.

In order to obtain a theory with $n$ fundamental hypermultiplets, one should 
add D4-branes parallel to the original $k$ D4-branes but located outside the interval 
in $x^6$ where the latter are located. These D4-branes end on either left or right NS5-branes and
extend to either $x^6=-\infty$ or $x^6=+\infty$. The two kinds of D4-branes will be called
left and right semi-infinite D4-branes, respectively, and their numbers denoted $n_L$ and $n_R$. 

The world-volume theory on the $k$ suspended D4-branes is now an $N=2$ $SU(k)$
gauge theory with $n_L+n_R$ hypermultiplets in the fundamental representation.
Their masses are given by $n_L+n_R$ complex numbers which parametrize the positions of
the semi-infinite D4-branes in the $45$ plane. Since the direction $3$ is periodic with
period $2\pi R$, the gauge theory is compactified on a circle of radius $R$.

In three dimensions the mass of the hypermultiplet is parametrized by three real numbers rather
than by one complex one. The same is true about the four-dimensional theory on a circle,
except that one of the three real mass parameters takes values in $\SS^1\cong \RR/\ZZ$ rather
than in $\RR$. Indeed, each hypermultiplet is associated with a global $U(1)$ symmetry.
Gauging this $U(1)$ symmetry and letting the Wilson line of the corresponding
photon along the compactified direction to be non-zero gives an effective mass to
the three-dimensional hypermuliplet. In the above string theory picture the global $U(1)$
is identified with the $U(1)$ gauge group of the semi-infinite D4-brane, and the extra
mass parameter is associated with the possibility of turning on a Wilson line along $x^3$
for the corresponding photon.

To interpret this brane configuration in terms of periodic monopoles, we perform T-duality along
$x^3$. The resulting configuration in Type IIB string theory consists of $k$ D3-branes suspended
between two NS5-branes, and $n_L+n_R$ semi-infinite D3-brane ending 
on the NS5-branes. 

\begin{center}
 \begin{tabular}{c|cccccccccc}
 &0&1&2&\raisebox{1pt}{$\bigcirc$}\hspace{-8.6pt}3&4&5&6&7&8&9\\
\hline
NS5 &x&x&x&x&x&x& & & & \\
D3&x&x&x& & & &x& & &
\end{tabular}
\end{center}

In the limit of $R\rightarrow 0$ the T-dual circle decompactifies and we end up with the
Chalmers-Hanany-Witten brane configuration. As pointed out in \cite{ChH}, in the worldvolume
theory on the two NS5-branes the suspended D3-branes appear as monopoles in the $SU(2)$
subgroup of $U(2)$, while the semiinfinite D3-branes appear as Dirac $U(1)$ monopoles
minimally embedded in $U(2)$. These monopoles live in the 
part of the NS5-branes' world-volume orthogonal to the D3-branes, namely in the 
$3,4,5$ directions. In the remaining directions ($0,1,2$) of the NS5-brane worldvolume 
the gauge field configuration on the NS5-branes is translationally-invariant.

For $R\neq 0$ the only difference is that the direction $3$ is compact. This means that
suspended D3-branes and semi-infinite D3-branes are nonabelian and Dirac monopoles on
$\RR^2\times\SS^1,$ respectively.
It can be checked that a left semi-infinite D3-brane corresponds to a Dirac monopole with $e_i=+1$,
while a right semi-infinite D3-brane corresponds to a Dirac monopole with
$e_i=-1$. Thus we can identify $n_L=n_+, n_R=n_-$.

As in~\cite{ChH}, one can argue that the metric on the moduli space of the suspended
D3-branes does not receive quantum corrections and thus is identical to the
classical metric on the moduli space of $k$ periodic monopoles with $n=n_L+n_R$
singularities. On the other hand, this same metric must be the metric on the Coulomb
branch of the $N=2$ $SU(k)$ gauge theory with $n$ fundamental hypermultiplets.

Note that the monopole charge $k$ is related to the rank of the gauge group of the $N=2$
gauge theory. If we require the $N=2$ gauge theory to be asymptotically free or finite,
$k$ must obey $2k\geq n$. In Section~\ref{sec:U2mon} we derived the same restriction
on the monopole charge by analyzing the asymptotic behavior of the periodic monopole. One may
wonder which assumption about monopoles corresponds to the requirement that the gauge theory
be asymptotically free or finite. The answer is quite simple: we assumed that the periodic monopole 
configuration breaks the gauge group $U(2)$ down to $U(1)\times U(1)$ for large $|z|$.
This is reflected in the fact that the difference of the eigenvalues of the Higgs field
either goes to infinity or approaches a finite non-zero limit for $|z|\ra\infty.$
That is, the assumption of asymptotic freedom or finiteness is equivalent to the
assumption of maximal symmetry breaking at infinity.

String theory picture can also be used to anticipate the result of Nahm transform applied to 
periodic monopoles. To this end we perform an S-duality and then a 
T-duality along the direction $3$. The resulting brane configuration
consists only of D4-branes.

\begin{center}
\begin{tabular}{c|cccccccccc}
 &0&1&2&\raisebox{1pt}{$\bigcirc$}\hspace{-8.6pt}3&4&5&6&7&8&9\\
\hline
D4&x&x&x& &x&x& & & & \\
D4&x&x&x&x& & &x& & &
\end{tabular}
\end{center}

Such a configuration of D4-branes is described by Hitchin equations on a cylinder parametrized by $x^3$ and 
$x^6$~\cite{SavAnton,Anton}.
The gauge group is $SU(K)$, where the number $K$ depends on $n_\pm$ and $k$. We will see below
that $K=\max(n_+,n_-,k)$.

\subsection{Periodic $U(m)$ Monopoles And $N=2$ Gauge Theories}

The relation between periodic $U(2)$ monopoles with and without singularities and quantum $N=2$ gauge theories
can be extended to $U(m)$ monopoles.  Consider the following gauge theory: the gauge group is
$SU(k_1)\times SU(k_2)\times\ldots\times SU(k_{m-1})$, the matter consists of $m-2$ hypermultiplets
in the representations 
$$
(k_1,\bk_2,1,\ldots,1), (1,k_2,\bk_3,1,\ldots,1), \ldots, (1,\ldots,1,k_{m-2},\bk_{m-1}),
$$ 
of the gauge group, $n_-$ hypermultiplets in the representation $(\bk_1,1,\ldots,1)$, and $n_+$ 
hypermultiplets
in the representation $(1,\ldots,1,k_{m-1})$. For $m=2$ this gauge theory reduces to the one studied
in the previous section. Its string theory realization consists of $m$ parallel NS5-branes along $0,1,2,3,4,5$
separated in the $x^6$ direction, $m-1$ stacks of D4-branes suspended between the successive NS5-branes
such that the $j^{\rm th}$ stack contains $k_j$ D4-branes, 
$n_-$ semi-infinite D4-branes ending on the right-most NS5-brane, and $n_+$ semi-infinite D4-branes
ending on the left-most NS5-brane~\cite{Witten}.

Performing T-duality along the $x^3$ direction converts all the D4-branes into D3-branes. The resulting
brane configuration in Type IIB string theory is identical to the one considered by Hanany and Witten~\cite{HW},
except that the direction $3$ is periodic. Since we have $m$ NS5-branes, their worldvolume theory
has gauge group $U(m)$.  The above brane configuration is represented in this worldvolume theory
by a BPS monopole with a nonabelian charge $(k_1,\ldots,k_{m-1})$ with $n_++n_-$
Dirac-type singularities of which $n_-$ have $e_i=-1$ and $n_+$ have $e_i=+1$~\cite{HW,usrthree3,usrthree1}.
We conclude that the moduli space of a $U(m)$ periodic monopole with $n_+$ singularities
with $e_i=+1$, $n_-$ singularities with $e_i=-1$ and a nonabelian charge $(k_1,\ldots,k_{m-1})$
is identical to the quantum Coulomb branch of the $N=2$ gauge theory described above compactified
on a circle.

Performing S-duality and then T-dualizing $x^3$ again yields a configuration consisting solely of
D4-branes. Such configurations have been studied in~\cite{SavAnton,Anton} in the case when the
direction $6$ is periodic and $k_1=k_2=\ldots=k_{m-1}$. The results of
\cite{SavAnton,Anton} suggest that brane configurations of this type are described by $SU(K)$
Hitchin equations on a cylinder, where $K$ is some integer number. As explained below,
Hitchin equations are related to periodic monopoles by means of the Nahm transform. The
integer $K$ will be determined to be $\max(n_+,n_-,k_1,\ldots,k_{m-1})$.

%%%%%%%%%%%%%%%%%%%%%%%%%%%%%%%%%%%%%%%%%%%%%%%%%%%%%%%%%%%%%%%%%%%%%%%%%%%

\section{Nahm Transform}
\label{Nahm}

This section describes the Nahm transform for periodic $U(2)$ monopoles with singularities, as well as
certain algebro-geometric data associated to monopoles. 

\subsection{Direct  Nahm Transform}
Given a periodic monopole $(A,\phi)$ with singularities
we define a family of Dirac-type
operators parametrized by a point $(r,t)\in\RR\times \RR/\ZZ$ as follows.
Let $L$ be a line bundle over $\RR^2\times\SS^1$ with a flat unitary connection 
$a=-t d\chi$ whose only non-zero component is along $\SS^1$. The variable $t$
takes values in the dual circle $\RR/\ZZ$ which we denote $\hSS^1$. Let $\sigma_i$ be 
Pauli matrices and let $r$ be a real number. Let $S$ be the trivial rank-two vector bundle on $\RR^2\times\SS^1$.
The Pauli matrices can be regarded as morphisms $S\ra S$. We define a first-order differential
operator $D:S\ot E\ra S\ot E$ by
\begin{equation}
D=\sigma\cdot d_{A+a}-(\phi-r).
\end{equation}
With some abuse of terminology we shall refer to it as ``the Dirac operator'', and to the 
bundle $S$ as ``the spin bundle''.

We shall use a multi-valued complex coordinate $s=r+it$ on $\RR\times\hSS^1$. 
It is important
to know for which values of $s$ the operator $D$ is Fredholm. If both $\ell_1$ 
and $\ell_2$ are non-zero, then it is easy to see that $D$ is Fredholm for all 
$s\in \RR\times\hSS^1$. If either $\ell_1$ or $\ell_2$ are zero, then one or both of 
the eigenvalues of the Higgs field stay finite for $|z|\ra\infty$.
Depending on whether $\ell_1=0$ or $\ell_2=0$, this eigenvalue is equal to $v_1$ or $v_2$. 
In this case $D$ can fail to be Fredholm only if $r=\rre s$ is equal to one of the 
finite eigenvalues, and at the same time $t=\iim s$ is equal to $b_1$ or $b_2$. 
This can be stated more concisely by introducing
a non-unitary connecton $A-i\phi d\chi$, and letting $V(z)$ to be its holonomy along the 
$\chi$ direction at a point $z\in\CC$. If both $\ell_1$ and $\ell_2$ are non-zero, 
the eigenvalues of $V(z)$ do not approach a finite limit as $z\ra\infty$; if $\ell_1=0$, 
$\ell_2\neq 0$, then one of the eigenvalues of $V(z)$ approaches
a finite limit $w_1=e^{v_1+ib_1}$; if $\ell_1\neq 0$, $\ell_2=0$, then one of the
eigenvalues of $V(z)$ approaches a finite limit $w_2=e^{v_2+ib_2}$; 
if $\ell_1=\ell_2=0$, then the eigenvalues of $V(z)$
approach $w_1$ and $w_2$. Let $\cK$ be the set of points on $\RR\times\hSS^1$ such that
$e^{2\pi s}$ is equal to one of the limiting eigenvalues of $V(z)$.
$\cK$ consists of at most two distinct points. 
The operator $D$ is Fredholm for $s\in (\RR\times\hSS^1)\bsl\cK$ because $D^\dag D$ has a mass gap.

The Weitzenbock formula~\cite{usone} implies that the operator $\Dd D$ is positive-definite on the 
space of functions of rapid decrease. It is also easy to see that all elements of the 
$L^2$ kernel of $D$ must be decreasing rapidly at infinity for $s\notin \cK$, and therefore 
the $L^2$ kernel of $D$ is empty for $s\notin\cK$.  It follows that the $L^2$ kernel of 
$\Dd$ is a vector bundle on  
$(\RR\times\hSS^1)\bsl\cK$ of rank $-\ind D$. We denote this bundle $\hE$. 
We will show below that for a periodic monopole
with singularities $\ind D=-\max(n_+,n_-,k)$. 

We now endow $\hE$ with a unitary connection $\hA$ and a section of $\End \hE$.
The space of all $L^2$ sections of $S\ot E$ forms a trivial unitary 
bundle over $\RR\times\hSS^1$ with a trivial connection. The bundle $\Ker\Dd$
is a subbundle in it. Let $P$ denote the orthogonal projector to $\Ker\Dd$. The induced connection
on $\Ker \Dd$ is given by 
\begin{equation}
\hA= i P ds\frac{\partial}{\partial s}+iPd\bs\frac{\partial}{\partial \bs}.
\end{equation}
The Higgs field $\hphi\in\Gamma(\End \hE)$ is defined by 
\begin{equation}
\phih(s)= P z .
\end{equation}
A computation similar to Nahm's original computation shows~\cite{usone} that $\hA$ and $\hphi$
satisfy the Hitchin equations:
\begin{eqnarray}
&&\bar{\partial}_{\hat{A}} \hat{\phi}=0\\
&& \hat{F}_{s\bar{s}}+\frac{i}{4}\left[\phih, \phih^{\dagger}\right]=0.
\end{eqnarray}
Thus to any periodic $U(2)$ monopole with singularities we can associate a solution of Hitchin equations on a
cylinder, with zero, one, or two points deleted, depending on the values of $\ell_1,\ell_2$.

\subsection{Reformulation Of The Nahm Transform}\label{sec:reform}
Here we present a useful reformulation of the Nahm transform in which one 
can easily recognize its cohomological origin. The cohomological formulation of the
Nahm transform is made explicit in~\cite{usone}. If we want to solve the Dirac equation 
\begin{equation}\label{Dir}
\Dd \theta=\Dd \begin{pmatrix}\theta_1\\ \theta_2\end{pmatrix}=0,
\end{equation}
we can do it in two steps. First, let us find a solution of the equation
\begin{equation}\label{cohom}
2(\partial_{\bar{z}}-i A_{\bar{z}})\uth_1-(\partial_{\chi}-\phi-i A_{\chi}+s)\uth_2=0,
\end{equation}
where $\uth_1$ and $\uth_2$ are $L^2$ sections of $E$.
If we define a first-order differential operator 
$$
\bD_1^s=\left(2(\partial_{\bar{z}}-i A_{\bar{z}}), 
-(\partial_{\chi}-\phi-i A_{\chi}+s)\right)
$$
which acts from $E\op E$ to $E$, and let 
$\uth=\uth_1\op\uth_2\in \Gamma(E\op E)$, then
the above equation can be rewritten as 
$$
\bD_1^s\uth=0.
$$ 

Now let us define an operator 
$\bD_0^s$ which acts from $E$ to $E\op E$:
\begin{equation}
\bD_0^s=\begin{pmatrix}\partial_{\chi}-\phi-i A_{\chi}+s \\ 
2(\partial_{\bar{z}}-i A_{\bar{z}})\end{pmatrix}.
\end{equation}
Bogomolny equations imply that $\bD_1^s \bD_0^s=0$. Thus if 
$\uth$ solves Eq.~(\ref{cohom}), 
then $\uth+\bD_0\rho$ also solves Eq.~(\ref{cohom}) for any $\rho\in \Gamma(E))$.

In the second step, we look for $\rho\in\Gamma(E)$ such that $\uth+\bD_0^s\rho$
is square-integrable and
solves the Dirac equation. Note that $\Dd$ can be written in the form
\begin{equation}
\Dd=\begin{pmatrix} \left(\bD_0^s\right)^\dagger\\ -\bD_1^s\end{pmatrix}.
\end{equation}
For
\begin{equation}
\theta=\uth+\bD_0^s\rho
\end{equation}
to solve $\Dd\theta=0$, the section $\rho$ must satisfy
\begin{equation}
\left(\bD_0^s\right)^\dagger\bD_0^s \rho=-\left(\bD_0^s\right)^\dagger \uth.
\end{equation}
Now observe that the operator $\left(\bD_0^s\right)^\dagger\bD_0^s$ is 
positive-definite for $s\notin\cK$, and therefore the above equation has a unique square-integrable
solution. Thus there is a one-to-one correspondence between solutions of the
Dirac equation $\Dd\theta=0$, and solutions of the equation
$\bD_1^s\uth=0$ modulo sections of $E\op E$ of the form $\bD_0^s\rho$.

The benefit of
the new description is that it simplifies the definition of $\hphi$ and
the holomorphic structure on $\hE$. Since the operators $\bD_1^s$ and $\bD_0^s$ 
commute with $\partial_s$ as well as with multiplication by a holomorphic function 
of $z$, the new definition of $\hphi$ is simply
$$
\hphi=z,
$$ 
and the new definition of the $\bpartial$ operator on $\hE$ is simply
$$
\bpartial_{\hA}=\frac{\partial}{\partial \bs}.
$$

\subsection{Monopole Spectral Data}
An important role in the subsequent analysis is played by the {\it spectral data}, which are 
algebro-geometric
data associated to every periodic monopole with singularities. To define the spectral data, 
consider the already familiar connection 
$B=A-i\phi d\chi$ and its holonomy $V(z)$ around the circle $\SS^1$ parametrized by $\chi$ at a point $z\in \CC\cong \RR^2$.
$V(z)$ is a section of a bundle obtained by restricting $E$ to the plane $\chi=0$. This bundle has a natural
holomorphic structure given by $\bpartial_A$, and the
Bogomolny equation ensures that $V(z)$ is a holomorphic section~\cite{usone}. Thus the coefficients of the
characteristic polynomial of $V(z)$ are holomorphic functions of $z$, and the equation
\begin{equation}\label{specM}
\det\left(e^{2\pi s}-V(z)\right)=0
\end{equation}
defines a holomorphic curve in $\CC\times \CC^*$, where we identified $\CC^*$ with the cylinder
$\RR\times\hSS^1$ parametrized by $s$ via the exponential map. This curve $\Sc$ will be called
the monopole spectral curve.
Since each point of the curve $\Sc$ corresponds to an eigenvalue of $V(z)$, there is a well-defined
sheaf $M$ on $\Sc$ consisting of the eigenvectors of $V(z)$. The stalk of $M$ at a general point is one-dimensional.
For a general monopole spectral curve is nonsingular, and the sheaf $M$ is a line bundle~\cite{usone}.
The line bundle $M$ has a natural holomorphic structure defined as follows. 
For $s$ and $z$ related by 
Eq.~(\ref{specM}), a section of $M$ is represented
by a section $\theta$ of the bundle $E$ satisfying
\begin{equation}\label{monod}
\left(\frac{\partial}{\partial\chi}-i A_{\chi}-\phi+s\right) \theta=0.
\end{equation}
It is a holomorphic section of $M$ if and only if
\begin{equation}
\left(\frac{\partial}{\partial \bar{z}}-i A_{\bar{z}}\right) \theta=0.
\end{equation}
These two equations are consistent because of Bogomolny equations.
We will call $M$ the spectral line bundle, and the pair $(\S,M)$ the monopole
spectral data.

\subsection{Hitchin Spectral Data}
To every solution of Hitchin equations on $\CC^*\cong \RR\times\hSS^1$ one can also 
associate a holomorphic curve $\C$ and a sheaf $N$ on it. The equation of the curve is 
the characteristic equation of $\hphi(s)$:
\begin{equation}\label{specN}
\det\left(z-\hphi(s)\right)=0.
\end{equation}
It defines a holomorphic curve in $\CC\times\CC^*$ because $\hphi$ is a holomorphic 
section of $\hE$ by virtue
of the Hitchin equations. The sheaf $N$ is the sheaf of eivenvectors of $\hphi(s)$. 
If the curve $\C$ is nonsingular,
then $N$ is a line bundle~\cite{usone}. The holomorphic structure on $N$ is defined as
follows: a section $\psi$ of $N$ is holomorphic if and only if
\begin{equation}
\left(\frac{\partial}{\partial {\bar{s}}}-i \hA_{\bar{s}}\right) \psi=0.
\end{equation} 
Since $\partial_{\bar{s}}-i \hA_{\bar{s}}$ commutes 
with $\hphi$, this definition is consistent. 
The pair $(\C,N)$ is called the Hitchin spectral data.

\subsection{Coincidence Of The Spectral Data}\label{sec:coin}
To any periodic monopole we can thus associate two kinds of spectral data: the monopole spectral
data, and the Hitchin spectral data of its Nahm transform. A fact of paramount importance is that these
two kinds of spectral data coincide. In~\cite{usone} we proved this for periodic monopoles without
singularities, but the argument applies to the present case just as well. Below we
sketch the construction of the isomorphism between the two kinds of spectral data.

Suppose that a point $(\zeta, e^{2\pi\sigma})$ belongs
to the spectral curve $\C\subset\CC\times\CC^*$. If $\Theta$ represents 
a holomorphic section of $N$, it satisfies
\begin{equation}\label{N1}
\hphi(\sigma) \Theta=\zeta\Theta,
\end{equation}
and
\begin{equation}\label{N2}
(\partial_{\bar{\sigma}}-i \hA_{\bar{\sigma}}) \Theta=0.
\end{equation}
{}From the point of view of Nahm transform, $\Theta$ is a zero mode of $\Dd$ 
twisted by $\sigma$. As explained above, we can also think of $\Theta$ as a
section $\uth\in \Gamma(E\op E)$ satisfying $\bD_1^\sigma\uth=0$, modulo
the equivalence relation $\uth\sim\uth+\bD_0^{\sigma}\rho$. 
{}From this point of view, Eq.~(\ref{N1}) is equivalent
to $(z-\zeta)\uth=\bD_0^{\sigma}\psi$. The latter equation implies that $\psi\vert_{z=\zeta}$
represents a section of the sheaf $M$ as defined in Eq.~(\ref{monod}). 
The statement that the holomorphic lines bundles $M$ and $N$ are isomorphic
means that the condition
\begin{equation}
\left(\frac{\partial}{\partial \bar{\zeta}}-i A_{\bar{\zeta}}\right) \psi\vert_{z=\zeta}=0
\end{equation}
on $\psi$
is equivalent to imposing the condition (\ref{N2}) on $\Theta$.

\subsection{Index Computation}

Let us now justify the claim that $\ind D=-\max(n_+,n_-,k)$. The operator
$D$ is a Callias-type, or Dirac-Schroedinger, operator. There
are a number of index theorems for these kind of operators which express $\ind D$
in terms of Chern classes of the eigenbundles of the Higgs field on the boundary. 
Unfortunately, none of this theorems applies to the present situation, as they 
usually do not allow for singularities of the fields. 

Instead of using
this direct approach, we will give two indirect
arguments which show that $\ind D=-\max(n_+,n_-,k)$.
The first argument uses the fact that the monopole spectral curve coincides
with the Hitchin spectral curve. By definition of the Nahm transform,
$-\ind D=\rank \hE$, which in turn is equal to the number of times the Hitchin spectral
curve covers the $w$-plane. On the other hand, it is shown in the next section
that the monopole spectral curve has the form
$$
A(z)w^2-B(z)w+C(z)=1,
$$
where $w=e^{2\pi s}$, and $A(z)$, $B(z)$, and $C(z)$ are polynomials of degree $n_-$, $k$, 
and $n_+$, respectively. This curve covers the $w$-plane $\max(n_+,n_-,k)$ times,
which proves our statement.

The second argument uses a reformulation of the Nahm transform 
presented in subsection~\ref{sec:reform}. It provides some information on the
spatial structure of the zero modes of $\Dd$.
Consider a complex of sheaves of vector spaces
$$
K^\sigma:\ 0\ra E\xrightarrow{\bD_0^\sigma} E\op E\xrightarrow{\bD_1^\sigma}
E\ra 0, 
$$ 
where the sections of $E$ are assumed to be rapidly decaying. The value of $\sigma$ is arbitrary,
except that we require $\sigma\notin\cK$. 
As shown in Section~\ref{sec:reform}, $H^1(K^\sigma)$ is naturally 
isomorphic to $\Ker \Dd$ twisted by $s=\sigma$.
It remains to compute the dimension of $H^1(K^\sigma)$. To this end consider 
another complex of sheaves:
$$
0\ra E\xrightarrow{(z-\zeta)} E\xrightarrow{rest.} E|_{z=\zeta}\ra 0,
$$
where $rest.$ is the restriction map, and $E|_{z=\zeta}$ is concentrated
on the circle $z=\zeta$.
This complex is not exact; nevertheless it leads to a long exact sequence in 
$\bD^\sigma$-cohomology~\cite{usone}:
$$
0\ra H^0_{\bD^\sigma}(\SS^1, E|_{z=\zeta})\ra H^1(K^\sigma)\xrightarrow{z-\zeta} 
H^1(K^\sigma)
\xrightarrow{rest.} H^1_{\bD^\sigma}(\SS^1, E|_{z=\zeta})\ra 0.
$$
This exact sequence implies that the dimension of $H^1(K^\sigma)$ is equal to the 
number of points at which 
the spectral curve intersects the line $s=\sigma$. 
Using the explicit form of the monopole spectral curve, one can easily see that
this number is equal to $\max(n_+,n_-,k)$.

\section{Boundary Conditions For Hitchin data}\label{sec:Hitchin}

\subsection{General Remarks}

The boundary conditions on the Hitchin data will be determined mainly by studying
the monopole spectral curve. It has the general form
$$
w^2-b(z)w+c(z)=0.
$$
The functions $b(z)=\Tr\ V(z,2\pi)$ and $c(z)=\det\ V(z,2\pi)$ are known to be
holomorphic on $\CC\backslash \{z_1,\ldots,z_n\}$. We now show that they are
rational functions.

Using the known asymptotic behavior of $\phi$ and $A$ near the singularities, 
we compute
that $c(z)$ has a simple zero at $z=z_i$ if $e_i=+1$ and a simple pole if $e_i=-1$.
As for $b(z)$, it has a simple pole at $z=z_i$ if $e_i=-1$ and is regular if $e_i=1$.
Thus $b(z)$ and $c(z)$ are meromorphic on $\CC$.

Using the known asymptotic behavior of $\phi$ and $A$ at infinity, 
we obtain the following asymptotic formulas for $b(z)$ and $c(z)$ for
$z\ra\infty$:
\begin{align}
b(z)&=z^{\ell_1}e^{\ups_1}\left(1+\frac{\mu_1}{z}+O(1/z^2)\right)+
z^{\ell_2}e^{\ups_2}\left(1+\frac{\mu_2}{z}+O(1/z^2)\right),\\
c(z)&=z^{\ell_1+\ell_2}e^{\ups_1+\ups_2}
\left(1+\frac{\mu_1+\mu_2}{z}+O(1/z^2)\right).
\end{align}
Here $\ups_1=v_1+ib_1,\ups_2=v_2+ib_2$. Hence $b(z)$ and $c(z)$ are meromorphic
on $\PP^{1}$, i.e. rational functions on $\CC$. Moreover, the above information about 
the poles of $b(z)$ and $c(z)$ implies
$$
b(z)=\frac{B(z)}{A(z)},\qquad c(z)=\frac{C(z)}{A(z)},
$$
where 
$$
A(z)=\prod_{e_i=-1}(z-z_i),\qquad C(z)=e^{\ups_1+\ups_2}\prod_{e_i=1}(z-z_i),
$$
and $B(z)$ is a polynomial of degree $k$.
Thus the monopole spectral curve can be rewritten in the following
form
\begin{equation}\label{curveSW}
A(z)w^2-B(z)w+C(z)=1.
\end{equation}
It is understood here that the points $(w,z)$ on the curve satisfying $A(z)=0$
must be deleted. 
It is important to note that the known asymptotics
of $b(z)$ and $c(z)$ determine the leading and the next-to-leading coefficients of
$B(z)$ in terms of the monopole parameters. For example, the leading coefficient
of $B(z)$ is given by
$$
\begin{cases}
e^{\ups_1}, &\ell_1>\ell_2\\
e^{\ups_1}+e^{\ups_2}, & \ell_1=\ell_2.
\end{cases}
$$
The precise expression for the next-to-leading coefficient will not be needed here.
The remaining $k-2$ coefficients of $B(z)$ are the moduli of the monopole.

In~\cite{Witten} it was shown that the curve~(\ref{curveSW}) is the 
Seiberg-Witten curve for the
$N=2$, $d=4$ gauge theory with gauge group $SU(k)$ and $n$ fundamental hypermultiplets.
The masses of the hypermultiplets are the zeros of $A(z)$ and $C(z)$, which
from the monopole point of view are just the positions of the Dirac-type 
singularities. The reason for this ``coincidence'' was explained in 
Section~\ref{gauge}.

We proved in Section~\ref{Nahm} that the Hitchin spectral curve is identical
to the monopole spectral curve. We now use the form of the spectral curve 
to determine the asymptotic behavior of the Hitchin data. The results depend
on the relative magnitude of the numbers $k,n_+,n_-$. There are
seven possible cases to consider. But note that the substitution $\phi\rightarrow-\phi$
$\ \chi\rightarrow-\chi$ leaves the Bogomolny equation invariant and in terms of 
the spectral curve maps $w$ to $1/w$ leaving $z$ invariant. Thus this map
interchanges $n_+$ and $n_-$, and without loss of generality we may assume that
$n_-\leq k$. This leaves us with four cases to consider.

\subsection{The case $n_-<k=n_+$}\label{firstcase}
The rank of $\hE$ is $n_+=k$. The set $\cK\subset \RR\times\hSS^1$ consists of a 
single point $w=w_2=e^{\ups_2}$. Thus we need to understand the behavior of $\hA$
and $\hphi$ for $|r|\ra\infty$, as well as near $w=w_2$.

We begin with the region of large $|r|$. The spectral curve equation
implies that the eigenvalues of $\hphi$ for $r\ra +\infty$ ($w\ra \infty$) asymptote to
\begin{multline}\label{Aas}
a_1, a_2, \ldots , a_{n_{-}}, 
e^{(2\pi s-\ups_1)/(k-n_{-})},\omega_{k-n_-} e^{(2\pi s-\ups_1)/(k-n_-)},\ldots,\\
\omega_{k-n_-}^{k-n_--1} e^{(2\pi s-\ups_1)/(k-n_-)},
\end{multline}
where $\omega_p=e^{2\pi i/p}$, and $a_1,\ldots,a_{n_-}$ are the roots of $A(z)$.
For $r\rightarrow-\infty$ ($w\ra 0$) the $n_+$ eigenvalues of $\hphi(s)$ are all 
distinct and approach the $n_+$ roots of $C(z)$.

To determine the behavior of $\hA$, note first that the curvature of $\hA$ goes
to zero for large $|r|$~\cite{usone}. Indeed, the computations in~\cite{usone} imply
that 
\begin{equation}\label{hF}
F_\hA=iP\sigma_3 (\Dd D)^{-1}P,
\end{equation}
where $P$ is the projector onto the kernel of $\Dd$. Now, it is easy to see that the
$L^2$ norm of the Green's function of $\Dd D$ is bounded from above by $const/|r|^{3/2}$,
and so is the norm of $F_\hA$. 
Since $F_\hA$ goes to zero at least as fast as $1/|r|^{3/2}$, $\hA$ 
has well-defined limiting holonomies around $\hSS^1$ for $r\ra \pm\infty$.

Since Hitchin equations relate $F_\hA$ and $[\hphi,\hphi^\dag]$,
we see that $[\hphi,\hphi^\dag]$ goes to zero for large $|r|$, and therefore in this limit
the eigenvectors of $\hphi$ become orthogonal.
Furthermore, since for $r\ra -\infty$ the eigenvalues of $\hphi$ approach
constants, in this limit the holonomy of $\hA$ becomes diagonal in the basis of the 
eigenvectors of $\hphi$. 

It is possible to find the eigenvalues
of the limiting holonomy by analyzing the Nahm transform in more detail.
Instead we will use a shortcut. As explained in the next section, the Nahm transform
admits an inverse. The inverse Nahm transform involves finding the kernel
of a family of Dirac-type operator $\hD^x$ parametrized by a point 
$x=(z,\chi)\in \RR^2\times\SS^1$. The fields $A(x),\phi(x)$ are expressed through the 
overlaps of the zero-modes of $\hD^x$. It will be seen that $\hD^x$ can fail to be 
Fredholm only when $\exp(i\chi)$ coincides with one of the eigenvalues of the
limiting holonomies of $\hA$, and $z$ coincides with the corresponding limiting
eigenvalue of $\hphi$. For all other values of $x=(z,\chi)$, the operator $\hD^x$ is 
Fredholm, and the fields $A(x),\phi(x)$ are nonsingular. We already know that the limiting 
eigenvalues of $\hphi$ for $r\ra -\infty$ are precisely the $z$-coordinates of the 
singularities with $e_i=+1$. Hence the limiting eigenvalues of $\hA$ must
be the $\chi$-coordinates of these singularities. Thus, if we denote these
$\chi$-coordinates by $\chi_{n_-+1},\ldots,\chi_n$, the holonomy of
$\hA$ in the basis of the eigenvectors of $\hphi$ and in the limit $r\ra -\infty$
is
\begin{equation}\label{rudder}
\diag(e^{i \chi_{n_-+1}}, e^{i \chi_{n_-+2}},\ldots, e^{i \chi_n}).
\end{equation}

Now let us find the limiting holonomy of $\hA$ for $r\ra +\infty$. Note that the
Hitchin equations together with the limiting behavior of the eigenvalues of
$\hphi$ imply that in the basis of the eigenvectors of $\hphi$
the limiting holonomy is
\begin{equation}
\begin{pmatrix} 
e^{i \beta_1}&      &                    &  \\ 
                   &\ldots&                    & \\ 
                   &      & e^{i \beta_{n_-}}& \\
                   &      &                    &V_{k-n_-} e^{i\talpha_1/(k-n_-)}
\end{pmatrix},
\end{equation}
where $\beta_1,\ldots,\beta_{n_-},\talpha_1\in \RR/(2\pi\ZZ),$
and $V_p$ is $p\times p$ ``shift'' matrix given by
\begin{equation}\label{shift}
\qquad V_{p}=
\begin{pmatrix}
0 & 1 & 0 & \dots & 0 & 0 \\
0 & 0 & 1 & \dots & 0 & 0 \\
0 & 0 & 0 & \dots & 0 & 0 \\
\hdotsfor{6} \\
0 & 0 & 0 & \dots & 0 & 1 \\
1 & 0 & 0 & \dots & 0 & 0 \\
\end{pmatrix}.
\end{equation}
The ``shift'' matrix appears because $k-n_-$ of the limiting
eigenvalues of $\hphi$ are cyclically permuted as one goes around $\hSS^1$. Note also that $\talpha_1$
takes values in $\RR/(2\pi\ZZ)$ because a shift $\talpha_1\ra\talpha_1+2\pi m, m\in\ZZ,$ can be undone 
by a gauge transformation.

It remains to determine $\beta_1,\ldots,\beta_{n_-}$
and $\talpha_1$. The first $n_+$ eigenvalues of the limiting holonomy are associated
with the eigenvalues of $\hphi$ which become constant in the limit $r\ra +\infty$.
Using the same shortcut as above, we find that $\beta_i$ is equal
to $\chi_i$, the $\chi$-coordinate of the $i^{\rm th}$ singularity with $e_i=-1$.
We will discuss how to express $\talpha_1$ in terms of the parameters of the monopole
in the end of this subsection.

Now let us determine the behavior of $\hA$ and $\hphi$ near the point $w=w_2$.
From the spectral curve equation it is easy to see
that the function $z(w)$ considered as 
a meromorphic function on the curve has a simple pole at $w=w_2=\exp(\ups_2)$ with 
residue $e^{\ups_2}\mu_2$. This implies that the Higgs field $\hphi$ behaves as
$$
\hphi(s)\sim \frac{R_1}{s-s_2}+O(1),
$$
where $s_2=\frac{1}{2\pi}\log w_2$, and $R$ is a rank-one matrix
whose non-zero eigenvalue is $e^{\ups_2}\mu_2$. 

Now let us show that $[R,R^\dag]=0$. To this end it is sufficient to demonstrate
that $[\hphi,\hphi^\dag]$ grows at most as $1/|s-s_2|$ in the limit $s\ra s_2$.
To estimate the norm of this commutator, note that by virtue of Hitchin equations
and (\ref{hF}) we have  
$$
[\hphi,\hphi^\dag]=-4 P\sigma_3 (\Dd D)^{-1} P.
$$
Now the required estimate on the commutator follows from a simple estimate
of the $L^2$ norm of $(\Dd D)^{-1}$.

Having determined the behavior of $\hphi$ near $s=s_2$, we now turn to $\hA$.
Our strategy will be the following. Specifying a unitary connection $\hA$
is equivalent to specifying a holomorphic structure on $\hE$, as well as
a Hermitian metric on $\hE$. In particular, the behavior of $\hA$ near $s=s_2$ is
determined by the rate of growth of holomorphic sections of $\hE$ near $s=s_2$.
As a basis of holomorphic sections we will use the eigenvectors of $\hphi$.
They can be reinterpreted as holomorphic sections of the spectral line bundle.
Their norm can be estimated using the ``cohomological'' reformulation of Nahm
transform described in Section~\ref{sec:reform}. The same strategy was used
in ~\cite{JB} to study the Nahm transform of instantons on $\RR^2\times T^2$,
and we will make use of some of the results of that paper.

An important role in this argument is played by the holomorphic isomorphism of
the spectral line bundles $M$ and $N$ whose construction we now recall.
Let $(\zeta, e^{2\pi\sigma})$ be the coordinates of
a point on the spectral curve $\C$. In the neighourhood of $\sigma=s_2$ 
$\zeta$ has a simple pole as a function
of $\sigma$, $\zeta\sim 1/(\sigma-s_2)$. If $\Theta$ is a vector in the fiber of the 
spectral line bundle over the point $(\zeta, e^{2\pi\sigma})$, then
$$\phih(\sigma)\Theta=\zeta\Theta.$$ 
The fact that $\zeta$ diverges for $\sigma\ra s_2$
means that $\Theta$ is an eigenvector of $\phi$, unique up to a multiple, 
whose eigenvalue diverges in this limit. Let $\theta$ be the corresponding class
in $\bD$-cohomology, as described in Section~\ref{sec:coin}. This implies
\begin{equation}\label{lull}
(z-\zeta)\theta=\bD_0^\sigma \psi_\zeta(z,\chi),
\end{equation}
for some $\psi_\zeta(z,\chi)\in\Gamma(E)$. Then $\psi_\zeta(\zeta,\chi)$ 
is an element in the fiber $M\ra\Sc$, and the isomorphism between $M$ and $N$ identifies
it with $\Theta$. Now let us start varying $\sigma$ so that $\Theta(\sigma)$ is
a holomorphic section of the spectral line bundle. This is equivalent to saying that
the section $\psi_\zeta(\zeta,\chi)$ is holomorphic with respect to $\bpartial_A$, i.e.
\begin{equation}\label{bow}
(\partial_{\bar{\zeta}}-i A_{\bar{\zeta}}) \psi_\zeta(\zeta,\chi)=0.
\end{equation}
{}From the asymptotics of $A$ we infer that when $\zeta$ tends to infinity
the norm of such a section behaves as $|\zeta|^{-n-\alpha_2/(2\pi)}$ with an 
integer $n$. 

Since $M$ and $N$ are holomorphically equivalent, the corresponding $\Theta$
is going to satisfy $(\partial_{\bar{\sigma}}-i \hA_{\bar{\sigma}}) \Theta(\sigma)=0$. 
For any $\mu(z,\chi)\in\Gamma(E)$ we can replace $\psi_\zeta(z)$ with $\psi_\zeta(z)+
(z-\zeta)\mu(z,\chi)$ without spoiling 
the holomorphicity condition~(\ref{bow}). If we define a representative
$\uth$ of the class $\Theta$ by $\uth_\zeta(z)=\bD_0\psi_\zeta(z)/(z-\zeta)$
such a change of $\psi_\zeta$ will change $\uth_\zeta$ by $\bD_0\mu$.
As explained in Section~\ref{sec:reform}, from $\uth_\zeta$ we can construct 
an $L^2$ solution of the Dirac equation $\Dd\theta_\sigma=0$:
\begin{equation}\label{arg1}
\theta_\sigma=\uth_\zeta-\bD_0^\sigma
\left(\left(\bD_0^\sigma\right)^\dagger\bD_0^\sigma \right)^{-1}
\left(\bD_0^\sigma\right)^\dagger\uth_\zeta,
\end{equation}
The norm of $\Theta_\sigma$ is defined to be the $L^2$ norm of $\theta_\sigma$.
Now we use Lemma 7.5 of~\cite{JB} to estimate this norm:
\begin{equation}\label{arg2}
|\bD_0^\sigma
\left(\left(\bD_0^\sigma\right)^\dagger\bD_0^\sigma \right)^{-1}
\left(\bD_0^\sigma\right)^\dagger\uth_\zeta|\leq 
\frac{{\rm const}}{|\sigma-s_2|} |\left(\bD_0\right)^\dagger \uth_\zeta |.
\end{equation}
Thus the norm of $\theta_\sigma$ is bounded from above by a multiple of
\mbox{$|\sigma-s_2|^{-1+n+\alpha_2/(2\pi)}$.}

In a similar manner one can estimate the norms of the holomorphic sections which correspond
to the eigenvalues of $\hphi$ which stay finite in the limit $s\ra s_2$. We find
that their norms remain bounded.

Now we are ready to determine the behavior of $\hA$ near the singularity.
Since the norms of all holomorphic sections grow not faster than powers of $|\sigma-s_2|$,
$\hA$ must have a simple pole at this point:
$$
\hA_s(\sigma)=\frac{Q}{\sigma-s_2}+O(1).
$$
The residue $Q$ must satisfy $[Q,Q^\dagger]=0$, for the same reasons as the residue
of $\hphi$. Furthermore, it must satisfy $[Q^\dagger,R]=0$ for the equation
$\bpartial_\hA\hphi=0$ to be satisfied. Thus it is possible to choose a basis in the fiber of $\hE$
over $s=s_2$ such that both $Q$ and $R$ are diagonal. The above estimates on the norm of 
holomorphic eigenvectors of $\hphi$ then imply that the eigenvalues of $Q$ restricted to 
$\Ker\ R$ are zero, and therefore $Q$ is a rank-one matrix. It is easy to see that by a 
gauge transformation one can make the single non-zero eigenvalue of $Q$ to be purely imaginary. We will
denote this purely imaginary number by $i\talpha_2/(4\pi)$. 
Further gauge transformations can shift $\talpha_2$ by multiples of $2\pi$, so
we should regard $\talpha_2$ as taking values in $\RR/(2\pi\ZZ)$.

It remains to understand how the parameters $\talpha_1$ and $\talpha_2$ depend
on the monopole parameters. Since $\Tr\ F_\hA =0$, we can obtain a relation
between various limiting holonomies of $\hA$ by integrating this equation over
$\hX$ and using the Stockes' theorem:
\begin{equation}\label{stockes}
\sum_{i=1}^n e_i\chi_i=\talpha_1+\talpha_2.
\end{equation}
This equation leaves the individual values of $\talpha_1$ and $\talpha_2$
undetermined. We conjecture that $\talpha_1=\alpha_1$ and $\talpha_2=\alpha_2$.
One piece of evidence in favor of this is that the relation~(\ref{stockes}) is
automatically satisfied due to~(\ref{shroud}). More compelling evidence is provided by our
estimate of the rate of growth of holomorphic sections of $\hE$ near the point $s=s_2$.
We found that the sections in the image of $Q$ are bounded from above by a multiple of 
$|s-s_2|^{-1+n+\alpha_1/(2\pi)},$ where $n$ is an integer. If this bound were
saturated, the equality $\alpha_1=\talpha_1$ would follow. Then $\talpha_2=\alpha_2$
would also follow by combining~(\ref{stockes}) and (\ref{shroud}).

\subsection{The case $n_-<k>n_+$}

The rank of $\hE$ is $k$. The set $\cK$ is empty, i.e. the Hitchin data are 
non-singular for all $s$. The fact that $\phi$ is non-singular can be seen from the
spectral curve equation
\begin{equation}\label{speccase}
B(z)-wA(z)-\frac{C(z)}{w}=0.
\end{equation}
If $k>n_\pm$, then the roots of this polynomial equation in $z$ have no singularities
as functions of $w\in\CC^*$.

The spectral curve equation implies that for $r\ra +\infty$ 
the asymptotics of the eigenvalues of $\hphi$ are given by Eq.(\ref{Aas}).
The asymptotic behavior of the eigenvalues of $\hphi$ for $r\ra -\infty$  is analogous:
\begin{multline}
c_1, c_2, \ldots , c_{n_{+}}, e^{(2\pi s-\ups_1)/(n_{+}-k)}, \\
\omega_{n_+-k} e^{(2\pi s-\ups_2)/(n_+-k)}, \ldots , 
\omega_{n_+-k}^{k-n_+-1} e^{(2\pi s-\ups_2)/(n_+-k)},
\end{multline}
where $c_1,\ldots,c_{n_+}$ are the roots of $C(z)$. Note that all
eigenvalues of $\hphi$ either approach a constant value, or grow exponentially 
with $|r|$.

Since the curvature of $\hA$ decays as $|r|^{-3/2}$ for large $|r|$, $\hA$
has well-defined limiting holonomies around $\hSS^1$. The same reasoning
as in the previous subsection implies that in the
basis of the eigenvectors of $\hphi$ the limiting holonomy for $r\ra +\infty$ must have
the form
\begin{equation}
\begin{pmatrix} 
e^{i \chi_{1}}&      &                    &                       \\ 
              &\ldots&                    &                       \\ 
              &      & e^{i \chi_{n_-}}&                         \\ 
              &      &                    &V_{k-n_-} e^{i\talpha_1/(k-n_-)}
\end{pmatrix},
\end{equation}
where $\chi_i, i=1,\ldots,n_-,$ are the $\chi$-coordinates of the singularites
with $e_i=-1$, and $\talpha_1\in \RR/(2\pi\ZZ).$ 
For $r\ra-\infty$ the limiting holonomy in the basis of the eigenvectors of
$\phi$ is given by
\begin{equation}
\begin{pmatrix} 
e^{i \chi_{n_- +1}}&      &                    &  \\ 
                   &\ldots&                    & \\ 
                   &      & e^{i \chi_{n_-+n_+}}& \\
                   &      &                    &V_{k-n_+} e^{-i\talpha_2/(k-n_+)}
\end{pmatrix},
\end{equation}
where $\chi_{n_-+1},\ldots,\chi_{n_-+n_+}$ are the $\chi$-coordinates of the singularities
with $e_i=+1$, and $\talpha_2\in\RR/(2\pi\ZZ).$ 

It remains to express
the parameters $\talpha_1,\talpha_2$ through the parameters of the monopole.
The Stockes' theorem again implies~(\ref{stockes}), but leaves the individual
values of $\talpha_1,\talpha_2$ undetermined. We conjecture that in fact
$\talpha_1=\alpha_1$ and $\talpha_2=\alpha_2$. The main evidence in favor of this
conjecture is the upper bound on the rate of growth of holomorphic sections for
$r\ra \pm\infty$. For example, consider the limit $r\ra +\infty$. The
bundle $\hE$ has a subbundle spanned by the eigenvectors of $\hphi$ with diverging
eigenvalues. The rank of this subbundle is $k-n_-$. Let $L$ be its top exterior
power. $L$ is a line bundle which inherits from $\hE$ a holomorphic structure, as well
as a Hermitian inner product. It is easy to see that holomorphic sections of this bundle
grow as $e^{r(2\pi n-\talpha_1)}$, where $n\in\ZZ,$ for $r\ra +\infty.$ On the other hand,
one can estimate the rate of growth using the coincidence of the
spectral data, as in subsection~\ref{firstcase}, and get that the holomorphic sections of $L$
are bounded by $e^{r(2\pi n-\alpha_1)}, n\in\ZZ.$ If the bound is saturated, then $\talpha_1=\alpha_1$,
and consequently $\talpha_2=\alpha_2.$

\subsection{The case $n_-<k<n_+$}
The rank of the Hitchin system is $n_+$. The set $\cK$ is empty, i.e. the Hitchin
data are defined everywhere on the cylinder. It follows from the spectral curve
equation that for $r\ra -\infty$ the eigenvalues of $\hphi$ approach
the $n_+$ roots of $C(z)$.
The same argument as in the case $n_-<k=n_+$ shows that the limiting holonomy of $\hA$
in the basis of the eigenvectors of $\hphi$ is given by Eq.~(\ref{rudder}). 

The spectral curve equation also implies that for $r\rightarrow +\infty$ $n_-$ 
of eigenvalues of $\hphi(s)$ approach $a_1,\ldots,a_{n_-}$,
$n_+-k$ of them asymptote to the $n_+-k$ roots of 
\begin{equation}
z^{n_+-k}=e^{(2\pi s-\ups_1)},
\end{equation}
and $k-n_-$ of them asymptote to $k-n_-$ roots of
\begin{equation}
z^{k-n_-}=e^{(2\pi s-\ups_2)}.
\end{equation}
The limiting holonomy of $\hA$ at $r\ra +\infty$ in the basis of the
eigenvectors of $\hphi$ is given by
\begin{equation}\label{limhol}
\begin{pmatrix} 
e^{i \chi_{n_-+1}}&      &                    &                      & \\ 
                &\ldots&                    &                      & \\ 
                &      & e^{i \chi_{n_-+n_+}}&                      &   \\ 
                &      &                    &V_{n_+-k} e^{-i\talpha_1/(n_+-k)}& \\
                &      &                    &                      &V_{k-n_-} e^{-i\talpha_2/(k-n_-)}
\end{pmatrix},
\end{equation}
where $\talpha_1,\talpha_2\in \RR/(2\pi\ZZ).$ Stockes' theorem again yields~(\ref{stockes}).
We conjecture that $\talpha_1=\alpha_1,\talpha_2=\alpha_2$, for the same reason as in the previous case.

\subsection{The case $n_-=k=n_+$}
The bundle $\hE$ has rank $k$. The set $\cK$ consists of two points given by
$w=w_1=e^{\ups_1}$ and $w=w_2=e^{\ups_2}$. For $r\ra +\infty$ the eigenvalues of $\hphi(s)$
approach the roots of $A(z)$, while for $r\ra -\infty$ they approach the roots of $C(z)$.
The limiting holonomies of $\hA$ are well-defined and are given by (\ref{rudder}) for 
$r\ra -\infty$ and by
$$
\diag(e^{i \chi_{1}}, e^{i \chi_{2}},\ldots, e^{i \chi_{n_-}}).
$$
for $r\ra +\infty.$

The analysis of the singularities at $w=w_{1,2}$ is the same as in the case $n_-<k=n_+$.
For either of the singular points one of the eigenvalues of $\hphi$ has a simple
pole. The residue is equal to $\mu_1 e^{\ups_1}$ for $w=w_1$ and to
$\mu_2 e^{\ups_2}$ for $w=w_2$. Together with the estimate $||F_\hA||\leq const/|w-w_i|$,
this implies that $\hphi$ behaves as
$$
\hphi\sim \frac{R_i}{s-s_i}, \quad s_i=\frac{1}{2\pi} \log w_i, \quad i=1,2,
$$ 
where $R_i$ is a rank-one matrix whose only non-zero eigenvalue is $\mu_i e^{\ups_i}$,
and which satisfies $[R_i,R_i^\dag]=0$. As for the connection, similar arguments
show that it behaves as
$$
\hA_s\sim \frac{Q_i}{s-s_i},
$$
where $Q_i$ is a rank-one matrix satisfying $[Q_i^\dag,R_i]=0$. The only non-zero
eigenvalue of $Q_i$ can be made purely imaginary by a gauge transformation and
will be denoted $i\talpha_i/(4\pi).$ Stockes' theorem implies the relation~(\ref{stockes}),
as before. The estimate of the rate of growth of the holomorphic sections of $\hE$
near the points $s=s_i$ suggests that in fact $\talpha_i=\alpha_i, i=1,2.$

In the case $n_-=k=n_+$ Nahm transform for periodic monopoles resembles very much 
Nahm transform for doubly-periodic instantons studied in~\cite{Jardim}. 
Recall that doubly-periodic instantons are solutions
of the $U(2)$ self-duality equation on $\RR^2\times T^2$ with finite action and 
vanishing first Chern class. Their Nahm transform is described by solutions of 
the Hitchin equations on a torus with two punctures~\cite{SavAnton,Anton,Jardim}.
The behavior of $\hA$ and $\hphi$
at the punctures is the same as above. The rank of the Hitchin bundle is given by the
second Chern class of the instanton bundle on $\RR^2\times T^2$.

Nahm data for periodic monopoles can be regarded as a limiting case of Nahm data for
doubly-periodic instantons.
The cylinder $\RR\times\hSS^1$ can be regarded as a degeneration
of the torus of~\cite{Jardim}. For example, if the torus is realized as a quotient of $\CC$
by the lattice generated by $1$ and $\tau$, one can consider the limit $\iim \tau\ra +\infty$.
The positions of the punctures should be held fixed in this limit. 
The nonabelian monopole charge $k$ corresponds to the 
instanton number. 

One can see the reason for this 
by looking at the monopole side of the story. The case $k=n_+=n_-$ is special
in that the eigenvalues of the monopole Higgs field $\phi$ approach constants at infinity.
Now recall that the Bogomolny equation is a reduction of the self-duality equation
to three dimensions. Thus a periodic monopole can be regarded as an instanton
on $Y=\RR^2\times\SS_\chi^1\times\SS_\theta^1$ invariant with respect to the translations of the
circle $\SS_\theta^1$. The relation between the self-dual connection $\tA$ on $Y$ and the
monopole fields on $X=\RR^2\times\SS^1_\chi$ is given by
$$
\tA=\psi^*(A)+\psi^*(\phi) d\theta,
$$
where $\psi:Y\ra X$ is the natural projection. The connection $\tA$ is self-dual
for any $k$ and $n_\pm$, but the case $k=n_+=n_-$ is special, because only in this
case the large-$|z|$ behavior of $\tA$ is that of a doubly-periodic instanton as defined
in~\cite{Jardim}. Of course, unlike in~\cite{Jardim}, our $\tA$ has singularities
for $z=z_i,\chi=\chi_i$. 

The origin of these singularities can be understood by analyzing how the limiting procedure
described above affects the instantons.
A doubly-periodic $U(2)$ instanton with charge $1$ can be regarded as made of
two ``monopole'' constituents. Each constituent has a fixed size, so its moduli describe
its position on $\RR^2\times T^2$. Thus a charge $1$ instanton has 8-dimensional moduli
space. (This interpretation of an instanton as a combination of two monopoles also arises
for calorons, i.e. instantons on $\RR^3\times\SS^1$, see~\cite{calorons} for details.)
The sizes of the two constituents are determined by the asymptotic behavior of the components
of $A$ along the $T^2$ and need not be the same. In particular, one can take a limit
in which the size of one of the constituents goes to zero, while the size of the other
one stays finite. A point-like consitituent monopole is nothing but a Dirac-type singularity
on $\RR^2\times T^2$ of the kind considered in this paper. Thus for $k=n_-=n_+$
periodic monopoles with singularities can be obtained as a limit of doubly-periodic
$U(2)$ instantons with instanton charge $k$.

The degeneration of a doubly-periodic instanton into a periodic monopole with singularities
can be easily seen by slightly modifying the brane configuration discussed in 
Section~\ref{gauge}. 
A doubly-periodic $U(2)$ instanton of charge $k$ is described by a brane configuration
with $k$ D3-branes and $2$ NS5-branes, but with the $x^6$ direction compactified on a circle. 
As described in Section~\ref{gauge}, a D3-brane can end on the NS5-brane, and therefore each 
D3-brane breaks into two segments suspended between the NS5-branes and capable of 
moving independently. These suspended segments represent the consituent monopoles.
The size of a constituent is inversely proporional to the length of the segment.
In order to obtain a charge $k$ periodic monopole with $2k$ singularities one has to
take the limit in which the NS5-branes are very close together, so that half of the
segments are much longer than the other half.

\section{Inverse Nahm Transform}
In this section we construct and study the inverse Nahm transform which associates a 
periodic monopole with singularities to a solution of Hitchin equations on
$(\RR\times\hSS^1)\bsl\cK$ with the spectral curve and asymptotic behavior 
described in Section~\ref{sec:Hitchin}.

\subsection{Construction Of The Monopole Fields}\label{InvNahm}

Let $\hA,\hphi$ be a solution of Hitchin equations with $\rank(E)=K$. We will
use these data to define a family of Dirac-type operators parametrized by a point
$(z,\chi)\in\CC\times\SS^1$. Let $\ha=-\chi dt$ be a unitary connection on a 
trivial line bundle on $\RR\times\hSS^1$. We define a Dirac type-operator
from $\hE\op \hE$ to $\hE\op \hE$ by
\begin{equation}\label{hDirac}
\hD=\begin{pmatrix}
-\hphi+z & 2\partial_{\hA+\ha}\\ 
2\bar{\partial}_{\hA+\ha}& -\hphi^{\dagger}+\bar{z}
\end{pmatrix}.
\end{equation}
Now let us assume that $\hA$ and $\hphi$ satisfy the kind of boundary
conditions described in Section~\ref{sec:Hitchin}. Standard arguments
show that $\hD$ can fail to be Fredholm only if $z$ is equal to one of the asymptotic
eigenvalues of $\hphi$ and $e^{i\chi}$ is equal to the corresponding eigenvalue of
the limiting holonomy of $\hA$. (We remind that the boundary conditions of 
Section~\ref{sec:Hitchin} imply that the holonomy of $\hA$ in the limit $|r|\ra\infty$
preserves all the eigenvectors of $\hphi$ corresponding to finite limiting eigenvalues
and permutes the rest of the eigenvectors.)
Let us denote the set of such points of $\CC\times\SS^1$ by $\cM$. Obviously, $\cM$
a finite set whose cardinality does not exceed $2K$. 

The Weitzenbock formula~\cite{usone} implies that the $L^2$ kernel of $\hD$ is trivial, therefore
$\Ker\hDd$ is a vector bundle on $\RR^2\times\SS^1$ of rank $-\ind \hD$. It is a 
subbundle of a trivial infinite-dimensional bundle on $\RR^2\times\SS^1$ whose
fiber consists of all $L^2$ sections of $\hE\op \hE$. The latter bundle has a natural
Hermitian inner product. Let $\hP$ be the corresponding projector to $\Ker\hDd$.
We define a connection $A$ on $E=\Ker\hDd$ by
\begin{equation}
d_A=\hP \left(dz\frac{\partial}{\partial z}+d\bar{z}\frac{\partial}{\partial \bar{z}}+
d\chi\frac{\partial}{\partial\chi}\right) \hP.
\end{equation}
We define a Higgs field $\phi\in \Gamma(\End(E))$ by
\begin{equation}
\phi=\hP r \hP
\end{equation}
Obviously, $\phi^\dag=\phi$.
These formulas are well-defined because the elements of the $L^2$ kernel of
$\hDd$ decay at least exponentially for $(z,\chi)\notin \cM$, as can be easily
verified.

We claim that $A,\phi$ satisfy the Bogomolny equation. The computation demonstrating
this is exactly the same as in~\cite{usone}. The proof that the
composition of the direct and inverse Nahm transform is the identity is also
the same as in~\cite{usone}. 

We now want to show that the solution of the Bogomolny equation obtained in this
way is a periodic $U(2)$ monopole with singularities in the sense of 
Section~\ref{sec:U2mon}. This will imply that there is a one-to-one correspondence
between periodic $U(2)$ monopoles with singularities and Hitchin data of the kind described
in Section~\ref{sec:Hitchin}.
As a first step, we need to show that $-\ind\hD=2$, and therefore the monopole bundle 
obtained by the inverse Nahm transform has the rank two.
The argument for this is exactly the same as in~\cite{usone}. One proves
that $\dim\Ker\hDd$ for $z=z_0,\chi=\chi_0$ is equal to the number of points at which 
the Hitchin spectral curve intersects the line $z=z_0$ (provided $(z_0,\chi_0)\notin\cM$).
On the other hand, one of our assumptions about the Hitchin system was that 
the Hitchin spectral curve has the form
$$
A(z)w^2-B(z)w+C(z)=0,
$$
where $w\neq 0$, and the roots of $A(z)$ and $C(z)$ are the asymptotic
eigenvalues of $\hphi$. It follows that $\dim\Ker\hDd=2$ for all $(z,\chi)\notin\cM$.

\subsection{Asymptotic Behavior Of The Monopole Fields}

It remains to show that the monopole has the right behavior near
the points of $\cM$, as well as for $|z|\ra\infty$.
Suppose that the boundary conditions for the Hitchin data are such that
$n_-$ of the eigenvalues of $\hphi$ approach constants as $r\rightarrow+\infty$
and $n_+$ of them approach constants as $r\rightarrow-\infty$.
Taking into account the $w\rightarrow 1/w$ symmetry we may assume that $n_-\leq n_+$
without loss of generality.
Recall now that we assumed that the spectral curve defined by
the equation
\begin{equation}
\det(z-\hphi(s))=0
\end{equation}
has the form
\begin{equation}
A(z)w^2-B(z)w+C(z)=0,
\end{equation}
where $w=e^{2\pi s}$, and $A(z)$ is a polynomial of degree $n_-$, $C(z)$ 
is a polynomial of degree $n_+$. If we denote by $k$ the degree of $B(z)$,
then it is easy to see that the rank of the Hitchin system $K$ is $\max(n_+,n_-,k)$. 
Moreover, analyzing the four
possible boundary conditions for $\hphi$, one can easily see that
$2k\geq n_++n_-$, and if one normalizes the leading coefficient of $A(z)$ to be
one, then the leading coefficient of $C(z)$ is
$$
e^{\ups_1+\ups_2}
$$
and the leading coefficient of $B(z)$ is
$$
\begin{cases}
e^{\ups_1}, &\ell_1>\ell_2,\\
e^{\ups_1}+e^{\ups_2}, & \ell_1=\ell_2,
\end{cases}
$$
where $\ell_1=k-n_-,\ell_2=n_+-k$.

We now use these properties of the spectral curve to find the
behavior of $\phi$ and $A$ for large $|z|$ as well as near the points of $\cM$.

To analyze the behavior for $|z|\ra\infty$, note the following formula
for the components of curvature of $A$~\cite{usone}:
\begin{equation}\label{FP}
F_{\bz\chi}=2i\hP\sigma_-(\hDd\hD)^{-1}\hP,\qquad F_{z\bz}=i\hP\sigma_3(\hDd\hD)^{-1}\hP.
\end{equation}
It is easy to see that the $L^2$ norm of $(\hDd \hD)^{-1}$ is bounded from
above by a multiple of $1/|z|$. Hence the components of curvature decay at least
as fast as that, and then the Bogomolny equation implies that the covariant
differential of $\phi$ goes to zero for $|z|\ra\infty$. It follows that for
large $|z|$ the eigenvalues of $\phi$ are independent of $\chi$,
and furthermore that the eigenvalues of $V(z)$ factorize into the eigenvalues
of the holonomy of $A$ and the eigenvalues of $e^{2\pi\phi(z,\chi)}$. Thus the
behavior of the eigenvalues of $\phi$ for large $|z|$ can be read off the asymptotic
behavior of the eigenvalues of $V(z)$, which are encoded in the spectral curve.
One can easily see that this yields (\ref{infHiggs}) with $\ell_1=k-n_-$
and $\ell_2=n_+-k$. 

The behavior of $A$ for large $|z|$ can be inferred from Theorem 10.5
of~\cite{JT} about the behavior of solutions of the Bogomolny equation.
(There the theorem is stated for the Bogomolny equation on $\RR^3$, but the proof goes 
through for $\RR^2\times\SS^1$ as well).
This theorem asserts that if the difference between the
eigenvalues of the Higgs field is bounded from below for $|z|\ra\infty$, then
$d_A\phi$ is proportional to $\phi$ with exponential accuracy. Thus, up to
corrections of order $\exp(-\delta|z|)$, $\delta>0$, the connection
$A$ preserves the splitting of $E$ into the eigenbundles of $\phi$. Hence if we
use the eigenvectors of $\phi$ as the orthonormal frame of $E$, the $U(2)$
Bogomolny equation splits into a pair of decoupled $U(1)$ Bogomolny equations.
Solving the rank-one Bogomolny equation for $A$ we find that the asymptotic behavior of 
$A$ is given by~(\ref{infA}) with some $\alpha_1,\alpha_2$. One can also check that
(\ref{derinfHiggs}) is satisfied.

We now turn to the behavior of $\phi$ and $A$ near the points of $\cM$.
Recall that their $z$-coordinates are given by the roots of $A(z)$ and $C(z)$,
while their $\chi$ coordinates are given by the corresponding limiting eigenvalues
of the holonomy of $\hA$. From the spectral curve equation we see that near such a
point $z=z_i$ one of the eigenvalues of $V(z)$ either diverges as $(z-z_i)^{-1}$, or 
goes to zero as $z-z_i$. Obviously, this can happen only if the Higgs field
$\phi$ becomes singular at $z=z_i,\chi=\chi_i$.

To determine the nature of the singularity, we first estimate how fast $F_A$ and $\phi$
can grow near the singular point. It is easy to show that
the norm of $(\hDd\hD)^{-1}$ is bounded by a multiple of $1/r_i^2$, where
$r_i$ is the distance to the singular point. By~(\ref{FP}), the same is true
about the norm of $F_A$. Further, one can show that for sufficiently small $r_i$ all 
the elements of $\Ker \hDd$ are bounded by a multiple of $\exp(-|r|r_i)$. (This
is due to the exponential decay of the Green's function $(\hDd\hD)^{-1}.$) Then
from the definition of $\phi$ one can easily see that
$$
||\phi||\leq \frac{const}{r_i}.
$$

In the appendix we prove that any solution of the $U(2)$ Bogomolny equation 
with such a singularity has the following form:
\begin{align}\label{singHiggsgen}
\phi(x)&\sim h(x)\begin{pmatrix} m_1\phi_0(r_i) & 0 \\ 0 & m_2\phi_0(r) \end{pmatrix}
h(x)^{-1}+O(1),\\ \label{singderHiggsgen}
d_A\phi(x)&\sim h(x)\begin{pmatrix} m_1 d\phi_0(r_i) & 0 \\ 0 & m_2 d\phi_0(r) \end{pmatrix}
h(x)^{-1}+O(1),\\ \label{singAgen}
A(x)&\sim h(x) \begin{pmatrix} m_1 A_0(x-x_i) & 0 \\ 0 & m_2 A_0(x-x_i)\end{pmatrix}
h(x)^{-1}\nn \\ 
& + ih(x) dh(x)^{-1}+O(1).
\end{align}
Here $h(x)$ is a $U(2)$-valued function defined in the neigborhood of the singular
point, $m_1,m_2$ are integers, and $\phi_0(r)$ and $A_0(x)$ have been defined in 
Section~\ref{sec:U2mon}.
The idea of the proof is to
lift the monopole to an instanton on the Taub-NUT space with a point deleted, use the
Uhlenbeck compactification theorem to show that the instanton can be continued
to the deleted point, and then project the instanton back to three dimensions.

Using these formulas for $A$ and $\phi$ it is straighforward to compute $V(z)$ near a 
singular point and compare with what the spectral curve predicts. The results
match if and only if one of the $m_i$ is zero, and the other one is $1$ or
$-1$, depending on whether $z_i$ is a root of $A(z)$ or $C(z)$.
This completes the demonstration that the inverse Nahm transform produces
a periodic monopole with singularities out of any solution of Hitchin equations
with the boundary conditions as in Section~\ref{sec:Hitchin}.

\section{Nahm Transform For Periodic $U(m)$ Monopoles}

Periodic $U(m)$ monopoles with singularities defined in 
Section~\ref{PUM} can be analyzed in the same manner as $U(2)$ monopoles.
Let us describe the result of the Nahm transform applied to a periodic $U(m)$
monopole with $n_+$ (resp. $n_-$) singularities of positive (resp. negative) Chern 
class and nonabelian charges $(k_1, k_2,\ldots, k_{m-1})$.

Nahm transform yields a solution of Hitchin equations on a cylinder with several
points deleted. The rank of the Hitchin bundle is equal to $\max(n_+,n_-,k_1,\ldots,
k_{m-1})$. The deleted points are determined as follows. Recall that the integers
$\ell_1,\ldots,\ell_m$ which determine the behavior of the eigenvalues of $\phi$
at infinity are given by
\begin{eqnarray}
l_1&=&k_1-n_+\nn\\
l_2&=&k_2-k_1\nn\\
&&\ldots\\
l_{m-1}&=&k_{m-1}-k_{m-2}\nn\\
l_m&=&n_- -k_{m-1}.\nn
\end{eqnarray}
These numbers are ordered: $\ell_1\geq \ell_2\geq\ldots\geq \ell_m$. If $\ell_i=0$,
then the $i^{\rm th}$ eigenvalue  of $\phi$ approaches a finite value $v_i$ 
for $|z|\ra\infty$, and the corresponding eigenvalue of the holonomy of $A$ along 
$\SS^1$ approaches a constant value $e^{b_i}$. Set $\ups_i=v_i+ib_i$. 
The deleted points are in one-to-one
correspondence with $i$ such that $\ell_i=0$, and they are located at 
$w=w_i=e^{\ups_i}$.

The Higgs field $\hphi$ has a simple pole at each of the deleted points:
$$
\hphi\sim\frac{R_i}{s-s_i}.
$$
Here $s_i=\ups_i/(2\pi)$ and $R_i$ is a rank-one matrix satisfying $[R_i,R_i^\dag]=0$ whose 
only non-zero eigenvalue is equal to $\mu_i w_i$ (see~(\ref{infHiggsM}) for the definition
of $\mu_i$). The connection $\hA$ also has a simple pole at the deleted point:
$$
\hA\sim\frac{Q_i}{s-s_i},
$$
where $Q_i$ is a rank-one matrix satisfying $[Q_i^\dag,R_i]=0$.

Before we describe the behavior of the Hitchin data for $r\ra \pm\infty$,
let us note two alternative definitions of the rank $K$ in terms of $\ell_i$.
Let $j_+$ be the number of strictly positive $\ell_i$. It is easy to see that
\begin{equation}\label{identone}
K=n_-+\sum_{i=1}^{j_+} \ell_i.
\end{equation}
Similarly, if $j_-$ is the number of strictly negative $\ell_i$, then
we have an identity
\begin{equation}\label{identtwo}
K=n_+-\sum_{i=m}^{m-j_-+1} \ell_i.
\end{equation}

The curvature of $\hA$ goes to zero as $const/|r|^{3/2}$ for $|r|\ra\infty$, therefore
$\hA$ has well-defined limiting holonomies for $r\ra \pm\infty$.

For $r\ra +\infty$ $n_-$ of the eigenvalues of $\hphi$ approach constant values
$$
z_1,z_2,\ldots,z_{n_-},
$$
where $z_i$ is the $z$-coordinate of the $i^{\rm th}$ singularity with $e_i=-1$.
The corresponding eigenvectors are also the eigenvectors of the limiting holonomy
of $\hA$ with eigenvalues
$$
e^{i\chi_1},e^{i\chi_2},\ldots,e^{i\chi_{n_-}},
$$
where $\chi_i$ is the $\chi$-coordinate of the $i^{\rm th}$ singularity with $e_i=-1$.

For $r\ra +\infty$ $\ell_1$ of the eigenvalues of $\hphi$ asymptote to
$$
\omega_{\ell_1}^j \exp\left(\frac{2\pi s-\ups_1}{\ell_1}\right), \quad j=1,\ldots,\ell_1,
$$
where $\omega_p$ denotes $\exp(2\pi i/p)$. The limiting holonomy of $\hA$ preserves the
subspace spanned by the corresponding eigenvectors and its restriction to this subspace
is equal to
$$
V_{\ell_1} \exp\left(2\pi i\alpha_1/\ell_1\right),
$$
where $V_{\ell_1}$ is the $\ell_1\times\ell_1$ ``shift matrix'' defined in~(\ref{shift}).
Further, for $r\ra +\infty$ $\ell_2$ of the eigenvalues of $\hphi$ approach
$$
\omega_{\ell_2}^j \exp\left(\frac{2\pi s-\ups_2}{\ell_2}\right), \quad j=1,\ldots,\ell_2,
$$ 
and so on, until we reach $\ell_{j_+}$ eigenvalues of $\hphi$ which approach
$$
\omega_{\ell_{j_+}}^j \exp\left(\frac{2\pi s-\ups_{j_+}}{\ell_{j_+}}\right), 
\quad j=1,\ldots,\ell_{j_+}.
$$ 
By~(\ref{identone}), we have described the behavior of all $K$ eigenvalues of $\hphi$ and
the limiting holonomy of $\hA$. Note that since all the numbers $\ell_1,\ldots,\ell_{j_+}$
are positive, the eigenvalues of $\hphi$ which do not approach finite values grow
exponentially as $r\ra +\infty$.
 
For $r\ra -\infty$ the situation is similar.
$n_+$ of the eigenvalues of $\hphi$ approach constant values
$$
z_{n_-+1},z_{n_-+2},\ldots,z_{n_-+n_+},
$$
where $z_{n_-+i}$ is the $z$-coordinate of the $i^{\rm th}$ singularity with $e_i=+1$.
The corresponding eigenvectors are also the eigenvectors of the limiting holonomy
of $\hA$ with eigenvalues
$$
e^{i\chi_{n_-+1}},e^{i\chi_{n_-+2}},\ldots,e^{i\chi_{n_-+n_+}},
$$
where $\chi_{n_-+i}$ is the $\chi$-coordinate of the $i^{\rm th}$ singularity with $e_i=+1$.
$|\ell_m|$ of the eigenvalues of $\hphi$ asymptote to 
$$
\omega_{\ell_m}^j \exp\left(\frac{2\pi s-\ups_m}{\ell_m}\right), \quad j=1,\ldots,|\ell_m|.
$$
The limiting holonomy of $\hA$ preserves the
subspace spanned by the corresponding eigenvectors and its restriction to this subspace
is equal to
$$
V_{\ell_m} \exp\left(2\pi i\alpha_m/\ell_m\right).
$$
$|\ell_{m-1}|$ eigenvalues of $\hphi$ asymptote to
$$
\omega_{\ell_{m-1}}^j \exp\left(\frac{2\pi s-\ups_{m-1}}{\ell_{m-1}}\right), 
\quad j=1,\ldots,|\ell_{m-1}|,
$$
and so on, until we reach the $|\ell_{m-j_-+1}|$ eigenvalues of $\hphi$ which
asymptote to
$$
\omega_{\ell_{m-j_-+1}}^j \exp\left(\frac{2\pi s-\ups_{m-j_-+1}}{\ell_{m-j_-+1}}\right), 
\quad j=1,\ldots,|\ell_{m-j_-+1}|.
$$ 
By~(\ref{identtwo}), we have described the behavior of all $K$ eigenvalues of $\hphi$
and the limiting holonomy of $\hA$. Note that since all the numbers $\ell_m,\ldots,
\ell_{m-j_-+1}$
are negative, the eigenvalues of $\hphi$ which do not approach finite values grow
exponentially as $r\ra -\infty$.

The spectral curve corresponding to such a periodic monopole can be determined either
by computing the characteristic polynomial of $\hphi(w)$, or directly from the definition
of the monopole. The latter way is simpler and yields the following equation
in $\CC\times\CC^*$:
$$
A(z)w^m+B_1(z)w^{m-1}+B_2(z)w^{m-2}+\ldots+B_{m-1}(z)w+C(z)=0.
$$
Here 
$$
A(z)=\prod_{i=1}^{n_-} (z-z_i),\qquad C(z)=e^{\ups_1+\ldots+\ups_m} 
\prod_{i=n_-+1}^n (z-z_i),
$$
and $B_i$ is a polynomial of degree $k_i$ whose leading and next-to-leading coefficients
are determined by the asymptotics of the monopole fields. Note that this is precisely
the Seiberg-Witten curve for the $N=2$ $d=4$ gauge theory corresponding to our
periodic $U(m)$ monopole~\cite{Witten}.

\section{Concluding Remarks}

In this paper we have studied periodic monopole with singularities. 
From the physical point of view, they are of interest for several reasons. First,
as we have shown above, their moduli spaces can be used to ``solve'' $N=2$ $d=4$
gauge theories compactified on a circle. Although computing the metric on the moduli space
of periodic monopoles is hard, it is still an infinitely easier problem than
summing up an infinite number of instantons in a quantum field theory, including
monopole loops wrapping the compactified direction. At the moment it is not clear if any
of these metrics can be computed in a closed form, but it seems reasonably straightforward
to compute the asymptotic expansion far along the Coulomb branch. This problem will be
addressed in~\cite{ALG}.

Second, we have seen that the Nahm transform for periodic monopoles with singularities
is described by Hitchin equations on a cylinder, and that the Hitchin spectral curve 
is the Seiberg-Witten curve of the corresponding gauge theory. Recall now
that the space of solutions of Hitchin equations is an algebraically completely integrable
system, $\hphi$ being the Lax operator, and $s$ being the spectral 
parameter~\cite{Hit,Mark}. 
Thus our approach allows to associate to any 
$N=2$ $d=4$ gauge theory which admits a brane realization an integrable system, so
that the Seiberg-Witten curve is given by the characteristic polynomial of the Lax
operator. The relation between $N=2$ $d=4$ gauge theories and completely integrable
systems was noted previously~\cite{Gorsketal,MW,DW}, but its origin remained somewhat
mysterious. For finite $N=2$ gauge theories this question was clarified 
in~\cite{SavAnton,Anton} (see also~\cite{Guk}). In our work on nonsingular
periodic monopoles~\cite{usone}, we described the Hitchin system corresponding to an $N=2$ 
super-Yang-Mills theory without matter, and in this paper we generalized this to
a much larger class of theories with product gauge groups and hypermultiplets.

Third, in Witten's approach to $N=2$ gauge theories~\cite{Witten}, the Seiberg-Witten curve
describes the world-volume of the M-theory fivebrane. Then BPS monopoles wrapping
the compactified direction are represented by Euclidean M2-branes whose boundaries lie
on the M5-brane. Computing the contribution of such configurations to the metric
on the moduli space seems rather difficult. On the other hand, our methods in principle
allow to compute the contribution of an arbitrary number of wrapped BPS monopoles.
Conceivably, this could shed some light on the dynamics of open M2-branes.

From the mathematical point of view, periodic monopole with singularities are of interest
because their moduli spaces provide new examples of hyperk\"ahler manifolds without
any continuous isometries. The interpretation of the moduli spaces in terms of $N=2$ 
gauge theories makes it clear
that they are complete (because the Higgs branch is absent) and 
locally flat at infinity (because the gauge theories are asymptotically free or finite).
In the case when the rank of the gauge group is one, the moduli space is an elliptic
fibration over $\CC$. No examples of complete hyperk\"ahler 
metric on elliptic fibrations which are locally flat at infinity were known prior to 
this work (a noncomplete example is
provided by the so-called Ooguri-Vafa metric~\cite{OV}), and it is an interesting question 
which fibrations admit such metrics. We will suggest a possible answer to this question
in~\cite{ALG}.

\section{Acknowledgments}
It is our pleasure to thank Nigel Hitchin, Marcos Jardim, and Tony Pantev for discussions.
S.Ch. is grateful to the Institute for Advanced Study, Princeton, for hospitality during the 
final stage of this work. S.Ch. was supported in part by NSF grant PHY9819686. 
A.K. was supported in part by DOE grant DE-FG02-90ER40542.

\appendix
\section{Singularities of the monopole fields}

Let $U\subset\RR^3$ be a punctured neighborhood of the origin, and let
$E$ be a rank two unitary vector bundle on $U$. Let $A$ be
a unitary connection on $E$ and $\phi$ be a Hermitian section of $\End(E)$ such that
the Bogomolny equation $F_A=*d_A\phi$ is satisfied and 
\begin{align}\label{bound}
||\phi||&\leq \frac{C_1}{|x|}, \\
||F_A||&\leq \frac{C_2}{|x|^2},
\end{align}
where $C_1,C_2>0$. We are going to show that there exist integers $m_1,m_2$
and a $U(2)$-valued function $h(x)$ defined on $U$ such that
\begin{align}\label{singHiggsapp}
\phi(x)&\sim h(x)\begin{pmatrix} m_1\phi_0(r) & 0 \\ 0 & m_2\phi_0(r) \end{pmatrix}
h(x)^{-1}+O(1),\\ \label{singAapp}
A(x)&\sim h(x) \begin{pmatrix} m_1 A_0(x) & 0 \\ 0 & m_2 A_0(x)\end{pmatrix}
h(x)^{-1}+ ih(x) dh(x)^{-1}+O(1).
\end{align}
Here $r=|x|$, $\phi_0(r)=-1/(2r),$ and $A_0(x)$ is defined in Section~\ref{sec:U2mon}.

Consider a four-dimensional noncompact manifold $X$ with coordinates $(x,t), x\in U,
\theta\in \RR/(4\pi\ZZ)$ and the metric 
\begin{align}\label{TNmetric}
ds^2=V(x) dx^i dx^j \delta_{ij}+V(x)^{-1} (d\theta+\omega_i(x) dx^i)^2,
\end{align}
where 
$$
V=1+\frac{1}{|x|},
$$
and $\omega=\omega_i(x) dx^i$ is a unitary connection on a line bundle on $U$ satisfying
the $U(1)$ Bogomolny equation:
$$
d\omega=*dV.
$$
It is easy to see that $\omega(x)=-A_0(x)$, and therefore the degree of the line bundle
is $-1$. 

The metric~(\ref{TNmetric}) is of the Taub-NUT (or Gibbons-Hawking) type,
and it is well known that it is hyperk\"ahler and admits a tri-holomorphic $U(1)$
action generated by the vector field $\frac{\partial}{\partial\theta}$.
The manifold $X$ endowed with this metric admits a nonsingular partial completion obtained by adding a single 
point $p$ over $x=0$. This point is invariant with respect to the $U(1)$ action mentioned
above. We denote this partial completion by $\bX$. 

Let $\psi$ be the projection $X\ra U$. Let $\tA$ be a connection on $\psi^*(E)$
defined by
\begin{equation}\label{tA}
\tA=\psi^*(A)+\psi^*\left(V^{-1}\phi\right)(d\theta+\omega).
\end{equation}
It is easy to check that it is a self-dual $U(1)$-invariant
connection on $\psi^*(E)$. The self-duality holds because $A$ and $\phi$ 
satisfy the Bogomolny equation. (This observation is due to P.~Kronheimer~\cite{Kron}.)
We claim that both $\psi^*(E)$ and $\tA$ can be continued in 
a unique manner to the point $p$ while preserving $U(1)$-invariance and self-duality.
Indeed, the action density of $\tA$ is given by
\begin{multline}
\Tr\ F_\tA \wedge *F_\tA=2\psi^*\left(V^{-2}\Tr\ F_A\wedge *F_A
+dV^{-1}\wedge *dV^{-1}\Tr\ \phi^2\right) d\theta\\
 - d\left[\psi^*\left(*dV^{-1} \Tr\ \phi^2\right)
d\theta\right].
\end{multline}
Using~(\ref{bound}), one can easily see that the integral of the
action density over $X$ converges. 
By Uhlenbeck's compactification
theorem~\cite{Uhl}, there is a unique smooth continuation of $\psi^*(E)$ and $\tA$ to $p$. 

Obviously,
the resulting connection is invariant with respect to $\frac{\partial}{\partial\theta}$ and
self-dual. It is easy to see that in a $\theta$-invariant gauge
any such connection has the following form
in the neighborhood of $p$:
$$
\tA=h(x)\begin{pmatrix} \left(\frac{m_1}{2}+O(x)\right)d\theta & 0 \\ 0 & 
\left(\frac{m_2}{2}+O(x)\right)d\theta \end{pmatrix} h(x)^{-1}
+ih(x) dh(x)^{-1}+a_i(x) dx^i.
$$ 
where $m_1,m_2$ are integers, $a_i(x)$ are smooth $u(2)$-valued functions,
and $h(x)$ is a smooth $U(2)$-valued function in the punctured neighborhood of $x=0$.
The geometric meaning of $m_1$ and $m_2$ is the following: they are the weights of
the $U(1)$ action on the fiber of $\psi^*(E)$ at $p$.
Solving~(\ref{tA}) for $A$ and $\phi$ and recalling that $\omega(x)=-A_0(x)$, we
obtain~(\ref{singHiggsapp}-\ref{singAapp}).

%%%%%%%%%%%%%%%%%%%%%%%%%%%%%%%%%%%%%%%%%%%%%%%%%%%%%%%%%%
 

\begin{thebibliography}{99}

\bibitem{AH}
M.~Atiyah and N.~Hitchin, ``The Geometry And Dynamics Of Magnetic Monopoles,''
{\it Princeton, USA: University Press (1988)}.

\bibitem{probe}
T.~Banks, M.R.~Douglas, and N.~Seiberg, ``Probing F Theory With Branes,''
Phys. Lett. {\bf B387}, 278 (1996) [hep-th/9605199].

\bibitem{JB}
O.~Biquard and M.~Jardim,
``Asymptotic behaviour and the moduli space of doubly-periodic  instantons,''
preprint math.dg/0005154.

\bibitem{ChH}
G.~Chalmers and A.~Hanany,
``Three dimensional gauge theories and monopoles,''
Nucl.\ Phys.\  {\bf B489}, 223 (1997)
[hep-th/9608105].

\bibitem{usone}
S.~A.~Cherkis and A.~Kapustin,
``Nahm transform for periodic monopoles and N = 2 super Yang-Mills  theory,''
preprint hep-th/0006050.


\bibitem{usrthree3} 
S.~A.~Cherkis and A.~Kapustin,
``Singular monopoles and gravitational instantons,''
Comm.\ Math.\ Phys.\  {\bf 203}, 713 (1999)
[hep-th/9803160].



\bibitem{usrthree2}
S.~A.~Cherkis and A.~Kapustin,
``D(k) gravitational instantons and Nahm equations,''
Adv.\ Theor.\ Math.\ Phys.\  {\bf 2}, 1287 (1999)
[hep-th/9803112].

\bibitem{usrthree1}
S.~A.~Cherkis and A.~Kapustin,
``Singular monopoles and supersymmetric gauge theories in three  dimensions,''
Nucl.\ Phys.\  {\bf B525}, 215 (1998)
[hep-th/9711145].


\bibitem{ALG}
S.~A.~Cherkis and A.~Kapustin,
``New Hyperk\"ahler Metrics From Periodic Monopoles,'' 
in preparation.


\bibitem{DW} R. Donagi and E. Witten, ``Supersymmetric Yang-Mills Theory And Integrable 
Systems,'' Nucl. Phys. {\bf B460}, 299 (1996) [hep-th/9510101].

\bibitem{GS}
O.~Ganor and S.~Sethi, ``New Perspectives On Yang-Mills Theories With Sixteen 
Supersymmetries,'' JHEP {\bf 9801}, 007 (1998) [hep-th/9712071].

\bibitem{Gorsketal} A.~Gorsky et al., ``Integrability And Seiberg-Witten Exact Solution,''
Phys. Lett. {\bf B355}, 466 (1995) [hep-th/9505035].

\bibitem{Guk}
S.~Gukov, ``Seiberg-Witten Solution From Matrix Theory,'' preprint hep-th/9709138.

\bibitem{HW}
A.~Hanany and E.~Witten,
``Type IIB superstrings, BPS monopoles, and three-dimensional gauge  dynamics,''
Nucl.\ Phys.\  {\bf B492}, 152 (1997)
[hep-th/9611230].

\bibitem{HitchinK3}
N.J.~Hitchin,
``Twistor construction of Einstein metrics,'' in: Global Riemannian Geometry (Durham, 1983),
p.~115, eds. T.J.~Willmore and N.J.~Hitchin, {\it Chichester: Horwood (1984).}

\bibitem{Hit}
N.~Hitchin, ``Stable Bundles And Integrable Systems,'' Duke Math. J. {\bf 54}, 91 (1987).

\bibitem{JT}
A.~Jaffe and C.~Taubes, ``Vortices And Monopoles. Structure Of Static Gauge Theories,''
{\it Boston: Birkh\"auser (1980)}.

\bibitem{Jardim}
M.~Jardim, ``Construction of doubly-periodic instantons,''  preprint math.dg/9909069;
``Nahm transform of doubly-periodic instantons,'' preprint math.dg/9910120;
``Spectral curves and the Nahm transform of doubly-periodic instantons,''
preprint math.ag/9909146.

\bibitem{Anton}
A.~Kapustin,
``Solution of N = 2 gauge theories via compactification to three  dimensions,''
Nucl.\ Phys.\  {\bf B534}, 531 (1998)
[hep-th/9804069].

\bibitem{SavAnton}
A.~Kapustin and S.~Sethi,
``The Higgs branch of impurity theories,''
Adv.\ Theor.\ Math.\ Phys.\  {\bf 2}, 571 (1998)
[hep-th/9804027].

\bibitem{geomeng}
A.~Klemm et al., ``Self-Dual Strings and $N=2$ Supersymmetric Field Theory,''
Nucl. Phys. {\bf B477}, 746 (1996)
[hep-th/9604034].

\bibitem{calorons}
T.C.~Kraan and P.~van~Baal,  ``Monopole Constituents Inside $SU(N)$ Calorons,''
Phys. Lett. {\bf B435}, 389 (1998) [hep-th/9806034].

\bibitem{Kron}
P.B.~Kronheimer, ``Monopoles and Taub-NUT metrics,'' M.Sc. Thesis, Oxford (1985).

\bibitem{Mark}
E.~Markman, ``Spectral Curves And Integrable Systems,'' Comp. Math. {\bf 93}, 255 (1994).

\bibitem{MW}
E.~Martinec and N.~Warner, ``Integrable Systems And Supersymmetric Gauge Theory,''
Nucl. Phys. {\bf B459}, 97 (1996) [hep-th/9509161].

\bibitem{OV}
H.~Ooguri and C.~Vafa, ``Summing Up D-Instantons,'' Phys. Rev. Lett. {\bf 77}, 3296
(1996) [hep-th/9608079].

\bibitem{SW}
N.~Seiberg and E.~Witten, ``Electric-Magnetic Duality, Monopole Condensation, And Confinement In
$N=2$ Supersymmetric Yang-Mills Theory,'' Nucl. Phys. {\bf B426}, 19 (1994), erratum ibid. {\bf B430},
485 (1994) [hep-th/9407087]; ``Monopoles, Duality, And Chiral Symmetry Breaking In $N=2$ Supersymmetric QCD,''
Nucl. Phys. {\bf B431}, 484 (1994) [hep-th/9408099].

\bibitem{SW3}
N.~Seiberg and E.~Witten, ``Gauge Dynamics And Compactification To Three Dimensions,''
in: The mathematical beauty of physics (Saclay, 1996), p. 333, 
{\it River Edge, NJ: World Sci. Publishing (1997)} [hep-th/9607163].

\bibitem{Simp}
C.T.~Simpson, ``Harmonic bundles on noncompact curves,'' J. Amer. Math. Soc. {\bf 3}, 713 (1990).

\bibitem{Uhl}
K.K.~Uhlenbeck, ``Removable singularities in Yang-Mills fields,'' Comm. Math. Phys. {\bf 83}, 11 (1982).

\bibitem{Witten}
E.~Witten,
``Solutions of four-dimensional field theories via M-theory,''
Nucl.\ Phys.\  {\bf B500}, 3 (1997)
[hep-th/9703166].


\end{thebibliography}
\end{document}